\documentclass[sigconf]{acmart}
\usepackage{popets}
\usepackage{tcolorbox}

\setcopyright{popets}
\copyrightyear{YYYY}

\acmYear{YYYY}
\acmVolume{YYYY}
\acmNumber{X}
\acmDOI{XXXXXXX.XXXXXXX}
\acmISBN{}
\acmConference{Proceedings on Privacy Enhancing Technologies }
\usepackage{tikz}
\usetikzlibrary{positioning}
\usetikzlibrary{shapes}
\usetikzlibrary{arrows.meta}
\usepackage{balance}

\usepackage{amsmath}   

\usepackage{listings}
\usepackage{xcolor}

\lstdefinestyle{jsonstyle}{
  basicstyle=\ttfamily\small,
  breaklines=true,
  columns=fullflexible,
  frame=single,
  keepspaces=true
}
\usepackage{multirow}

\usepackage[utf8]{inputenc}
\usepackage[T1]{fontenc}

\begin{document}

\title{Disclosure Divergence: Measuring Privacy Policy and Data Safety Misalignment at Scale}


\author{Mst Eshita Khatun}
\affiliation{%
  \institution{Louisiana State University}
  \city{Baton Rouge}
  \state{Louisiana}
  \country{USA}
  }
\email{mkhatu3@lsu.edu}

\author{Lamine Noureddine}
\affiliation{%
 \institution{Louisiana State University}
  \city{Baton Rouge}
  \state{Louisiana}
  \country{USA}
  }
\email{lnoureddine@lsu.edu}

\author{Sideeq Bello}
\affiliation{%
 \institution{Louisiana State University}
  \city{Baton Rouge}
  \state{Louisiana}
  \country{USA}
  }
\email{sbell49@lsu.edu}

\author{Aisha Ali-Gombe}
\affiliation{%
 \institution{Louisiana State University}
  \city{Baton Rouge}
  \state{Louisiana}
  \country{USA}
  }
\email{aaligombe@lsu.edu}






\renewcommand{\shortauthors}{Khatun et al.}

\begin{abstract}
With the rapid growth of mobile applications, user data privacy has become an increasing concern. While privacy policies describe how apps collect and share data, platforms such as Google Play provide Data Safety labels intended to summarize these practices. Because these disclosure channels are declared separately, they may present inconsistent representations of app data practices, creating uncertainty for users and regulators.
In this work, we conducted a large-scale empirical study of disclosure consistency across 6,051 Android apps. Using an LLM-based extraction framework and a unified schema over 14 Google Play data categories and two operations (collection and sharing), we measure per-app and per-category consistency and introduce a sensitivity-weighted risk score that emphasizes high-risk data types.
We find that misalignment disproportionately affects sensitive categories such as personal information and device identifiers, with sharing disclosures exhibiting lower consistency than collection disclosures. Elevated privac risk is concentrated in app categories associated with persistent monitoring and communication. Overall, our findings highlight structural gaps in current disclosure mechanisms and underscore the need for stronger verification and greater transparency in platform-level privacy reporting.
\end{abstract}


\maketitle
\section{Introduction}
The rapid growth of mobile applications has raised serious concerns about the privacy of user data \cite{kumar2022large, liu2021have, zimmeck2019maps, rodriguez2025privacy}. Mobile apps, in particular, have become deeply embedded in our everyday lives, continuously accessing and processing sensitive information such as location, contacts, identifiers, financial records, and health data. This pervasive data collection amplifies users’ exposure to tracking, profiling, and secondary data uses that often occur without explicit consent or understanding. In response to these growing privacy risks, Google introduced Data Safety Labels on the Play Store in 2022 \cite{google_datasafety_guide, khandelwal2024unpacking}, requiring developers to self-report what data their apps collect, share, and how such data are secured. Developers must also accompany these disclosures with privacy policies, which serve as formal, legally binding statements of an app’s data-handling practices and responsibilities \cite{android_datasafety_playintegrity}. Together, Data Safety Labels and privacy policies represent the main transparency interface between developers and users, informing users about how their data are collected, shared, and secured, and are thus ostensibly designed to enhance user trust when downloading and using apps.

Despite these efforts, persistent uncertainty remains over whether these declarations accurately reflect actual data practices or merely offer a perception of transparency that fails to guarantee real protection of users' privacy data. In particular, the lack of consistency between what developers self-report in their Data Safety Labels and what they state in their privacy policies about user data collection and sharing may confuse users, weaken their trust, and compromise the reliability and efficiency of Google’s overall privacy and transparency framework \cite{koch2022keeping, zhang2023privacy}.  

Prior studies have largely examined these two transparency layers in isolation. Research on policy analysis (e.g., PolicyLint \cite{andow2019policylint}, PrivOnto \cite{oltramari2018privonto}, Polisis \cite{harkous2018polisis},  Privacify \cite{woodring2024enhancing}, and PrivacyCheck \cite{nokhbeh2020privacycheck}) has focused on identifying data practices, contradictions, and compliance gaps within privacy policies. Meanwhile, work on Data Safety and Privacy Labels (e.g., Zhang et al. \cite{zhang2022usable, zhang2023privacy, zhang2024exploring}, Khandelwal et al. \cite{khandelwal2024unpacking, khandelwal2023comparing}, Khedkar et al. \cite{khedkar2024android} and Wang et al. \cite{wang2025big}) has evaluated their usability, coverage, and reliability as self-reported disclosures. 
In terms of cross-layer analysis, two recent studies have examined inconsistencies between platform privacy labels and privacy policies in both Android and iOS ecosystems~\cite{alomar2025effect,jain2023atlas}. These efforts provide important evidence that disclosure inconsistencies remain widespread across mobile platforms. However, several analytical dimensions remain underexplored. In particular, these prior works have given limited attention to how inconsistencies differ between data collection and data sharing practices, how disagreement patterns vary across data categories, and how the relative privacy sensitivity of different disclosure mismatches can be systematically characterized. As a result, while these existing studies establish the presence of disclosure inconsistencies, there remains a limited understanding of their structure, concentration, and potential privacy severity across the broader Google Play ecosystem.


Our findings reveal widespread inconsistencies between privacy policies and Google Play Data Safety Labels. At the disclosure level, approximately one third of collection (33\%) and sharing (31\%) disclosures disagree across the two layers, while over 90\% of apps contain at least one inconsistency in Data Safety Label disclosures. These discrepancies disproportionately affect sensitive categories such as Personal Information, Device Identifiers, Location, and Financial Data, and are primarily driven by omissions in Data Safety Labels relative to privacy policies. Importantly, an ablation analysis on generic sharing statements shows that the observed asymmetry between sharing and collection disclosures remains stable, reinforcing the robustness of our findings. The persistence of this asymmetry suggests that the discrepancies are not solely artifacts of broad policy language. Instead, they appear particularly pronounced in data-sharing disclosures and may partially reflect a service-provider loophole, where data practices described in privacy policies under third-party or service-provider contexts are not consistently represented in the corresponding Data Safety Labels. Moreover, our privacy risk scoring analysis shows that about half of the analyzed apps fall into a medium-risk tier and a small but non-trivial subset into a high-risk tier, indicating that disclosure misalignment remains pervasive even among highly rated and widely downloaded apps. 

In summary, this work makes the following salient contributions:
\begin{itemize}
    \item A large-scale measurement of 6,051 real-world Android apps spanning 14 Google Play data categories, quantifying inconsistencies in developers’ reported data-collection and data-sharing practices across Data Safety Labels and Privacy Policies.
    \item A framework that systematically and rigorously quantifies the consistency of developers’ privacy disclosures across Data Safety Labels and Privacy Policies, using a sensitivity-weighted risk scoring mechanism at both the app and data-category levels.
    \item Concrete insights and recommendations to help platform providers, regulators, auditors, and developers improve the accuracy and alignment of privacy disclosures, thereby enhancing user trust and engagement within the app ecosystem.
    \item To support reproducibility, we publicly release all artifacts associated with this study through a GitHub repository \cite{ArtifactsPPDDSL2026}
\end{itemize}


The remainder of this paper is organized as follows. Section \ref{motiv_backg} presents the background, motivation and research questions. Section \ref{methodology} describes the methodology, while Section \ref{results} reports the findings. Section \ref{discussion} discusses  additional qualitative analysis and key observations, recommendations, limitations, and future work. Section \ref{RW} reviews related work, and Section \ref{conclusion} concludes the paper.

\section{Background \& Motivation}
\label{motiv_backg}

Google Play’s Data Safety section requires developers to self-report which categories of user data their apps collect and share, and for what purposes, as shown in Appendix \ref{DSLEx}, Figure~\ref{fig:dsl}. Based on the developer documentation, the Google Play Data Safety schema defines 14 user data type categories—Location, Personal info, Financial info, Health and fitness, Messages, Photos and videos, Audio files, Files and docs, Calendar, Contacts, App activity, Web browsing, App info and performance, and Device or other IDs which together cover 38 specific data types such as Approximate location, Email address, Purchase history, Web browsing history, and Device or other IDs including others \cite{google_datasafety_guide}. The label, which distinguishes between “data collected” and “data shared with third parties,” is intended to give users a quick, standardized view of an app’s data practices directly in the store interface. In parallel, developers are also required (in most jurisdictions by the law, such as GDPR \cite{GDPR2016info} and CCPA \cite{CCPA2018}) to provide a privacy policy, typically hosted on an external website, that describes in more detail what data is processed, collected, and/or transferred; for which purposes; with which partners, and under what legal bases.

Together, these two artifacts form the primary interface for transparency between users and developers, where most users see the Data Safety label first and consult the privacy policy only if they have concerns. However, both artifacts are developer self-reported, and neither is systematically checked against the other or against the app’s actual behavior at scale. It therefore remains unclear how closely these descriptions agree, how they differ across data categories, and what these differences imply for user privacy. Our study addresses this by quantifying the degree of alignment and sensitivity risk of misalignment between Data Safety Labels and privacy policies across a large corpus of Google Play apps.

\subsection{Motivation}
Mobile apps increasingly mediate sensitive aspects of daily life, from communication and social networking to banking, health, and mental well-being. When deciding whether to install an app, users must primarily rely on two sources of information about data practices: the Google Play Data Safety label and the app’s privacy policy. If these disclosures are incomplete or inconsistent, users may receive an inaccurate or partial understanding of how their data is collected and shared, particularly concerning highly sensitive categories such as location, financial records, or health information.

This raises a practical question for both end users and platform governance: to what extent can current disclosure mechanisms be trusted to provide a consistent and reliable representation of an app’s stated data practices? Therefore, our goal in this paper is to provide a large-scale, quantitative assessment of the alignment between Data Safety labels and privacy policies, and to characterize the severity of any mismatches using a sensitivity privacy risk score. By systematically measuring consistency (agreement between the two transparency layers), identifying the data types with the highest misalignment prevalence, and assessing the sensitivity risk across thousands of apps, we aim to strengthen the reliability and accountability of privacy transparency mechanisms on Google Play. This study will help users make informed decisions, enable regulators to evaluate compliance, and guide platform designers in improving disclosure mechanisms and auditing practices. 
\subsection{Research Questions}
\label{rq}
To address these gaps, this study investigates inconsistencies, misalignment prevalence and sensitivity risk between privacy policies and Google Play’s Data Safety labels through the following research questions:

\begin{itemize}
    \item RQ1. How consistent are privacy policies and Google Play’s Data Safety labels when reporting data collection and sharing?
    \item RQ2. Which data categories are most frequently misaligned across the two disclosure layers?
    \item RQ3. How are apps distributed across low, medium, and high Sensitivity Risk Score tiers, and what does this distribution reveal about the severity of privacy misalignment between privacy policies and Data Safety labels?
    \item RQ4. Which app categories exhibit the highest privacy misalignment risk, and how is risk distributed within those categories?
\end{itemize}

Overall, these questions allow us to provide insight into how well privacy disclosures align in practice. By quantifying how often misalignment occurs, and which apps and data categories are most affected, our exploratory analysis highlights concrete risks for user privacy and helps guide more trustworthy transparency mechanisms.




\section{Methodology}
\label{methodology}
\begin{figure}[ht]
\centering
\includegraphics[width=0.5\textwidth, height= 5.9cm]{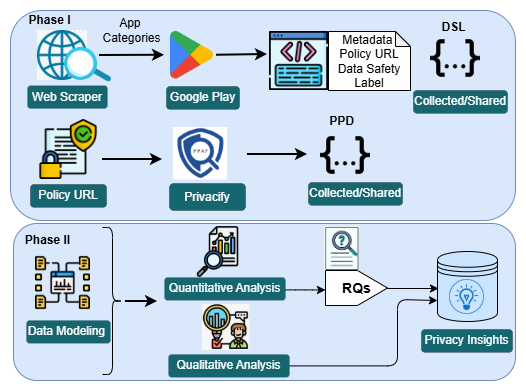}
\caption{Cross-layer privacy disclosure analysis pipeline}
\Description{methodology}
\label{fig:methodology}
\end{figure}

This section presents the methodology of our cross-layer privacy disclosure analysis pipeline for examining developers' privacy disclosures regarding user data across privacy policy documents and Data Safety Labels, as illustrated in Figure~\ref{fig:methodology}. Our methodology consists of two main phases: Phase I, data collection; Phase II, data modeling; followed by quantitative analysis and the derivation of insights related to user data privacy.

\subsection{Data Collection (Phase I)}\label{datacollection}
To investigate inconsistencies, the prevalence of misalignment, and the sensitivity risk of privacy disclosure, we first created a dataset of mobile applications from the Google Play Store. The data collection phase involved gathering three primary artifacts for each selected app: (1) the app’s privacy policy, (2) the Data Safety label displayed on its store listing, and (3) relevant metadata such as the app category, user ratings, and number of downloads. These artifacts serve as the foundational sources for evaluating our research questions. We developed a custom Python-based web-crawling toolkit targeting Google Play Store listings. The toolkit consists of modular scripts that (i) enumerate mobile applications from category-specific search queries, (ii) retrieve their Data Safety Labels, (iii) store the corresponding privacy policies, and (iv) collect basic app-level metadata (category, rating, reviews, downloads, and price). All components are implemented using \texttt{requests} and \texttt{BeautifulSoup4} for HTML fetching and parsing. Our crawler focuses on English-language apps, filtering out non-English listings during collection. This design yields a matched corpus of apps for which we have both Data Safety disclosures and privacy policies.

\subsubsection{Data Safety Labels}
We construct the Data Safety Label corpus starting from application category-specific search queries on the Google Play Store. For each query, our URL collection script crawls paginated search results, identifies individual app listing pages, and derives the corresponding Data Safety endpoints from their URLs. To avoid redundant work across runs, the crawler maintains a manifest of tuples of discovered Data Safety URLs and skips any URL that has already been processed. For each URL, a dedicated scraper fetches the Data Safety page and parses its HTML structure to locate the disclosure sections - \emph{Data collected} and \emph{Data shared}. It then iterates over the associated data categories, extracting both the sub-category label (data-type group) and its textual description. We normalize these disclosures into a structured JSON representation, where each app key maps to two sections (\emph{Data collected} and \emph{Data shared}), and each section is a mapping from a Data Safety category label (e.g., \emph{Location}, \emph{App activity}) to one or more concrete data-type strings (e.g., \emph{Approximate location}, \emph{App interactions}). For downstream analysis, we flatten this structure into a set of \emph{(category, data-type)} pairs per app. Before processing a new app, the scraper checks this JSON store; if an app is already present, it is skipped. This incremental design allows the pipeline to be re-run safely and produces a stable, reproducible corpus of Data Safety label annotations for our analysis, with a
representative JSON data structure illustrated in Appendix \ref{dslex} listing \ref{lst:bluesky-dsl}.

\subsubsection{Privacy Policy Documents}
To obtain privacy policies that correspond to the same set of apps, we leverage the Data Safety endpoints as stable entry points into each app’s detail page. A separate crawling component loads each app’s listing (or Data Safety) URL, locates the outbound \emph{Privacy policy} hyperlink in the app information panel, and records the resulting pair \emph{(app name, policy URL)} in a manifest file. As with the Data Safety crawler, we deduplicate entries by both app identifier and URL, ensuring that previously seen policies are not revisited in later runs.

Given this manifest, a policy downloader fetches the content of each privacy policy, preferring English-language variants when multiple localized versions are available. The HTML is sanitized to remove boilerplate and navigation elements and converted into a text format that preserves structural cues such as headings, lists, and tables. The resulting normalized policies serve as input to our LLM-based policy analysis pipeline, from which we later extract app developer practices regarding data type collection and sharing claims for comparison against the Data Safety Labels.

\subsubsection{Privacy Policy Analysis Pipeline} 
\label{backend validation}
For the analysis of privacy policy documents, our methodology relies on an LLM-based system to extract (i) the set of user data types an app claims to collect and (ii) the set of data types it reports sharing with third parties. To this end, we adopt and extend \textit{Privacify}~\cite{woodring2024enhancing}, a recent LLM-based privacy policy analysis system. Privacify implements a tokenization and chunking pipeline combined with an iterative map–reduce style summarization and recombination procedure, which enables processing long policies while preserving cross-section context. We further customize the final feature-extraction prompts so that the system outputs two sets of data types, \(C_{\text{data}}\) (collected) and \(S_{\text{data}}\) (shared with third parties), which we directly ingest as labels for our dataset construction.

\textbf{Validation Methodology.}
To assess extraction accuracy across diverse and representative conditions, we constructed a validation set of 100 apps using a stratified sampling strategy that covers app categories, popularity, and policy complexity.
\begin{itemize}  
\item For \textit{app category coverage}, we selected 60 apps ---  approximately 5 per category --- spanning 12 Google Play categories: Finance, Health and Fitness, Communication, Social, 
Shopping, Tools, Productivity, Education, Entertainment, Games, Lifestyle, and Personalization. This stratification ensures that category-specific vocabulary and disclosure patterns are represented in our evaluation, since privacy policy language varies substantially across domains. 
\item For \textit{app popularity}, we selected 20 apps based on install and rating signals: 5 low-install apps (approximately 1K--5K downloads), 5 high-install apps (10B+ downloads), 5 highly rated apps (rating = 5.0), and 5 low-rated apps (rating $\leq$ 2.0). 
\item For \textit{policy complexity}, we selected 20 apps based on structural and linguistic features: 5 very short policies (under 300 words), 5 very long policies (over 10,000 words), 5 policies with frequent third-party/SDK-related terms, and 5 policies using generic sharing language (e.g., “we may share,” “partners,” or “affiliates”). This strategy captures a range of policy structures and disclosure patterns to evaluate extraction performance under diverse privacy-policy conditions. 
\end{itemize}
For each of the 100 apps, one author manually annotated whether the policy indicates collection and/or sharing of each of the 14 data categories. In many cases, third-party data sharing was 
described only in generic terms (e.g., ``we may share your information''). When the policy did not specify which data categories were shared, we conservatively treated all data categories marked as collected for that app as shared with third parties. To assess the sensitivity of our data extraction results to this design choice, we further perform an ablation analysis under alternative treatments of generic sharing statements, described in Appendix \ref{ablation}. To further evaluate annotation reliability, a subset of 40 apps was independently annotated by a second annotator, yielding Cohen's $\kappa = 0.78$ and 90\% observed agreement. Disagreements were resolved through discussion before finalizing the ground truth labels.

\textbf{LLM Backend Selection and Parameter Configuration}. 
To assess whether extraction performance depends on the choice of LLM backend, we evaluated both the original Privacify backend, Mistral-7B, and our deployed backend, LLaMA 3.1 8B Instruct, on the same expanded 100-app validation set. For a fair comparison, both models were evaluated using the same preprocessing pipeline, chunking strategy, prompts, and output schema. Specifically, privacy policies were segmented using a token-based splitter with a chunk size of 3,000 tokens and an overlap of 100 tokens. Both models were run with deterministic decoding using temperature = 0.0, top-p = 1.0, and a maximum generation length of 6,000 tokens. The context window was configured to 8,500 tokens, and long policies were handled using chunking followed by map–reduce aggregation.

\begin{table}[ht]
\centering
\caption{LLM backend comparison on 100-app validation.}
\label{tab:llm-comparison}
\setlength{\tabcolsep}{4pt}
\begin{tabular}{llccc}
\toprule
\textbf{Operation} & \textbf{Model} & \textbf{Precision} & \textbf{Recall} & \textbf{F1-Score} \\
\midrule
\multirow{2}{*}{Collected}
& LLaMA 3.1 8B & 91.4\% & 93.0\% & 92.2\% \\
& Mistral-7B   & 90.5\% & 89.5\% & 90.0\% \\
\midrule
\multirow{2}{*}{Shared}
& LLaMA 3.1 8B & 89.9\% & 92.1\% & 91.0\% \\
& Mistral-7B   & 87.5\% & 88.1\% & 87.8\% \\
\bottomrule
\end{tabular}
\end{table}
Table~\ref{tab:llm-comparison} summarizes the extraction performance of the two LLM backends on the expanded 100-app validation set. LLaMA 3.1 8B Instruct consistently outperformed Mistral-7B across both collection and sharing extraction tasks. For collection disclosures, LLaMA 3.1 8B achieved an F1-score of 92.2\% compared to 90.0\% for Mistral-7B. The difference was more pronounced for sharing disclosures, where LLaMA 3.1 8B achieved an F1-score of 91.0\% versus 87.8\% for Mistral-7B. We also observed higher recall improvements for sharing extraction, suggesting that LLaMA 3.1 8B is more effective at identifying third-party sharing disclosures in privacy policies. Overall, these findings indicate that model choice has a measurable impact on extraction accuracy, particularly for sharing-related disclosures, while also demonstrating that the overall extraction pipeline remains stable across different LLM backends.  Overall, these results indicate that LLM extraction is sufficiently accurate to support aggregate measurements and provides a scalable, reliable mechanism for generating consistently labeled data for privacy policy analysis. An example of the resulting JSON data structure is provided in Appendix~\ref{ppdEx}, Listing~\ref{lst:bluesky-policy-datatypes}.

\color{black}
\subsubsection{Dataset Overview}
To construct the app corpus, we adopted a category-guided sampling strategy rather than relying on a single popularity ranking or top-chart list. Specifically, we defined category-specific search queries to cover a broad range of Google Play segments spanning diverse application domains and app genres. Apps were collected through query-based crawling across categories to better capture the heterogeneity of the Google Play ecosystem. For each query, our crawler retrieved app listing URLs from Google Play search results along with the corresponding Data Safety page, privacy-policy link, and available metadata.

Using this approach, we collected approximately 10K app entries from the Google Play Store between August and September 2025. We further removed duplicate records using the \texttt{APP Name} field and, when an app with a similar name appeared under multiple categories, we retained a single primary category. We also inspected the reported privacy policy URLs and discarded cases where the “policy” link actually pointed to unrelated websites (e.g., marketing pages or generic homepages) rather than a genuine privacy policy. This de-duplication and filtering step yielded a final corpus of 9{,}162 distinct applications spanning 419 application categories. These categories include generic genres such as \emph{Games}, \emph{Productivity}, and \emph{Lifestyle} and privacy-sensitive verticals, such as \emph{Health and Fitness}, \emph{Medical}, \emph{Mental health}, \emph{Finance}, \emph{Banking}, \emph{Insurance}, \emph{VPN apps}, \emph{Messaging}, \emph{Cloud storage}, \emph{Password manager}, and \emph{Kids education}. It also covers specialized app categories such as \emph{IoT control}, \emph{Smart home}, \emph{Remote desktop} and \emph{Wearables}. This range provides us with a diverse selection of applications, data types, and sensitivity levels to base our study on privacy disclosures. The summary of our dataset is shown in Table~\ref{tab:dataset-summary} including additional metrics such as free vs. paid , ratings, reviews and downloads.


\begin{table}[h]
\centering
\caption{Summary of the app corpus used in our analysis.}
\label{tab:dataset-summary}
\begin{tabular}{ll}
\hline
\textbf{Metric} & \textbf{Value} \\
\hline
Number of apps & 9{,}162 \\
Number of categories & 419 \\
App Rating & 4.19 (median 4.40)\\
Engagement (min-max) & 5 -- 212.4M (median 8.6K)\\
Apps with privacy policy & 6{,}051 (66.0\%) \\
Apps without privacy policy & 3{,}111 (34.0\%) \\
Free apps & 8{,}757 (95.6\%) \\
Paid apps & 405 (4.4\%) \\
Free apps with privacy policy & 5{,}910 (64.5\%) \\
Paid apps with privacy policy & 141 (1.5\%) \\
Free apps without privacy policy & 2{,}847 (31.1\%) \\
Paid apps without privacy policy & 264 (2.9\%) \\
Rating range (min--max) & 1.0 -- 5.0 \\
Reviews range (min--max) & 5 -- 212.4M \\
Downloads range (min--max) & 1K -- 500M \\
\hline
\end{tabular}
\end{table}

\subsection{Data Modeling (Phase II)}
After collecting the disclosure artifacts and metadata, the next phase involved modeling and structuring the dataset to enable comparison across privacy policies and Data Safety labels. We transformed the raw disclosures into standardized representations by parsing and categorizing data practices into a unified schema aligned with Google Play’s data categories and operations. This normalization process allows us to operationalize inconsistencies and misalignments at the data-type level and supports the computation of metrics such as misalignment prevalence and sensitivity-risk scores used in subsequent analysis.

We consider a set of apps indexed by 
\( i = 1,\dots,N \),
and a set \(C\) of data categories
(e.g., Location, Personal Info, Identifiers, Financial Info).
For each app \(i\), category \(c \in C\), and operation 
\( o \in \{\text{collect},\text{share}\} \),
we define binary indicators.

\subsection*{Binary Indicators}

The Privacy Policy Document (PPD) indicator:
\begin{equation}
PPD_{i,c}(o) =
\begin{cases}
1, & \text{if the privacy policy reports $c$ for $o$},\\
0, & \text{otherwise}.
\end{cases}
\end{equation}

The Data Safety Label (DSL) indicator:
\begin{equation}
DSL_{i,c}(o) =
\begin{cases}
1, & \text{if the Data Safety label reports $c$ for $o$},\\
0, & \text{otherwise}.
\end{cases}
\end{equation}

\subsection*{Consistency in the Privacy Policy Document and Data Safety Label}
We define consistency (alignment) as the agreement between the PPD and the DSL regarding whether a specific data category is collected or shared. This definition yields a binary consistency indicator for app $i$, category $c$, and operation $o$:

\begin{equation}
\mathrm{Cons}_{i,c}(o) =
\begin{cases}
1,&\text{if } PPD_{i,c}(o) = DSL_{i,c}(o)\\
0,& \text{otherwise.}
\end{cases}
\end{equation}

where $\mathrm{Cons}_{i,c}(o) = 1$ indicates a match (or an agreement) between the
privacy policy and the Data Safety label for category $c$ in operation $o$,
and $\mathrm{Cons}_{i,c}(o) = 0$ indicates a mismatch (misalignment or contradiction).

\subsection*{Consistency Score}

And thus, the aggregated app-level consistency score for an app $i$ and an operation $o$ is given as:
\begin{equation}
  \mathrm{ConsScore}_i(o)
  =
  \frac{1}{T_C}
  \sum_{c \in C}
  \mathrm{Cons}_{i,c}(o)
  \label{eq:consscore1}
\end{equation}
where $C$ denotes the set of data categories and $T_C = |C|$ is the total
number of categories.

\subsection*{Misalignment in the Privacy Policy}

Misalignment in the Privacy Policy Document is when the PPD omits but the DSL reports operation on data category. This is given as:
\begin{equation}
U^{PPD}_{i,c}(o) =
\begin{cases}
1, & \text{if } PPD_{i,c}(o)=0 \text{ and } DSL_{i,c}(o)=1,\\
0, & \text{otherwise}.
\end{cases}
\end{equation}

\subsection*{Misalignment in the Data Safety Label}

Misalignment in the Data Safety Label is when the DSL omits but the PPD reports operation on data category. This is given as:
\begin{equation}
U^{DSL}_{i,c}(o) =
\begin{cases}
1, & \text{if } PPD_{i,c}(o)=1 \text{ and } DSL_{i,c}(o)=0,\\
0, & \text{otherwise}.
\end{cases}
\end{equation}

\subsection*{Misalignment Score}
Thus, the misalignment indicator for an app $i$, category
$c$, and operation $o$ is given as misalignment in the Privacy Policy  \emph{or} misalignment in the Data Safety Label.

\begin{equation}
\mathrm{Mis}_{i,c}(o) =
\begin{cases}
1,& \text{if } U^{PPD}_{i,c}(o) = 1 ||\text{if }  U^{DSL}_{i,c}(o) = 1\\
0,& \text{otherwise.}
\end{cases}
\end{equation}

where $\mathrm{Mis}_{i,c}(o) = 1$ indicates a mismatch (contradiction or a disagreement) between the
privacy policy and the Data Safety label for category $c$ in operation $o$,
and $\mathrm{Mis}_{i,c}(o) = 0$ indicates a match (alignment).

And thus, the aggregated app-level misalignment score for an app $i$ and an operation $o$ is given as:
\begin{equation}
  \mathrm{MisScore}_i(o)
  =
  \frac{1}{T_C}
  \sum_{c \in C}
  \mathrm{Mis}_{i,c}(o)
  \label{eq:misscore}
\end{equation}
where $C$ denotes the set of data categories and $T_C = |C|$ is the total number of categories.

\subsection*{Consistency Vs Misalignment Scores}
\noindent Given the definition, Misalignment and Consistency form complementary measures over the unit interval, satisfying 
\begin{equation}
    \mathrm{ConsScore}_i(o) + \mathrm{MisScore}_i(o) = 1 
    \label{eq:consscore2}
\end{equation}







\subsection{Cohen's Kappa Score}
\label{sec:kappa}
The consistency metrics defined in Equations~\ref{eq:consscore1} and~\ref{eq:consscore2} capture observed agreement between privacy policies and Google Play’s Data Safety Labels; however, they do not adjust for agreement expected by chance. To address this limitation, we compute Cohen’s kappa ($\kappa$), a chance-corrected measure of inter-rater agreement, separately for the two disclosure operations: \textit{collection} and \textit{sharing}. We treat the PPD and the DSL as two \textit{raters}. For a fixed operation $o \in \{\text{collect}, \text{share}\}$, each \textit{app--category pair} $(i,c)$ constitutes one binary decision over the 14 Google data categories:
\[
PPD_{i,c}(o) \in \{0,1\}, \quad DSL_{i,c}(o) \in \{0,1\},
\]
where 1 indicates that the source discloses the category $c$ under operation $o$, and 0 indicates it does not.

Let $\Omega_o$ denote the set of evaluated app--category pairs for operation $o$, and let $N_o = |\Omega_o|$. The observed agreement ($P_o$) is the fraction of pairs where the two sources match:
\[
P_o(o) = \frac{1}{N_o} \sum_{(i,c)\in \Omega_o} \mathbb{I}\big[PPD_{i,c}(o)=DSL_{i,c}(o)\big].
\]
To correct for agreement that can occur due to the marginal tendency of each source to disclose (or not disclose) categories, we compute the expected agreement ($P_e$) from the marginal disclosure rates:
\[
p^{PPD}_1(o) = \frac{1}{N_o} \sum_{(i,c)\in \Omega_o} PPD_{i,c}(o), \quad
p^{DSL}_1(o) = \frac{1}{N_o} \sum_{(i,c)\in \Omega_o} DSL_{i,c}(o),
\]
with $p^{PPD}_0(o)=1-p^{PPD}_1(o)$ and $p^{DSL}_0(o)=1-p^{DSL}_1(o)$ 
\\The expected agreement is:
\[
P_e(o)=p^{PPD}_1(o)\,p^{DSL}_1(o)+p^{PPD}_0(o)\,p^{DSL}_0(o)
\]
Finally, Cohen's kappa is computed as:
\begin{equation}
    \kappa(o)=\frac{P_o(o)-P_e(o)}{1-P_e(o)}
    \label{eq:kappa}
\end{equation}

We report $\kappa$ score for collection and sharing to provide a chance-corrected measure of consistency between the two disclosure layers.
We use $\kappa$ in addition to percent agreement because both sources frequently report ``not disclosed'' across many categories, which can inflate raw agreement; $\kappa$ discounts agreement attributable to these marginal base rates.


\subsubsection{Sensitivity Risk Score (SRS)}
\label{sec:srs-definition}
Building on the consistency indicators above, we define a privacy-sensitive risk score that weights each misalignment by the sensitivity of the underlying data category. This metric is crucial because data categories differ substantially in their potential for privacy harm. Accordingly, we adopt a standard weighted aggregation approach for combining heterogeneous dimensions with unequal importance. The resulting $\mathrm{SRS}$ represents the expected misalignment intensity under a sensitivity-weighted distribution over data categories. For instance, a mismatch involving a device model has a fundamentally different risk profile than a mismatch involving financial credentials or health status. Treating all discrepancies equally obscures the severity of disclosure failures. Additionally, a simple count of inconsistencies treats all violations equally, preventing the distinction between “inconvenient” and “dangerous”. Therefore, weighting emphasizes the violations that matter most. Thus, our weighting misalignments metric, based on data sensitivity, produces a more realistic measure of disclosure risk than treating all inconsistencies equally.

Let $C$ denote the set of data categories and let $w_c \in \{w_1, w_2, \ldots, w_n\}$ be a sensitivity weight for category $c \in C$. For an app $i$, category $c$, and operation $o \in \{\text{collect},\text{share}\}$, we define the operation-specific SRS as:

\begin{equation}
\label{eq:srs-op}
\mathrm{SRS}^{(o)}_i
= \frac{\sum_{c \in C} w_c \bigl(\mathrm{MisScore}^{(o)}_{i,c}\bigr)}
       {\sum_{c \in C} w_c}.
\end{equation}

This resulting $\mathrm{SRS}^{(o)}_i$ represents a normalized sensitivity-weighted average of disclosure misalignment intensity across data categories for an operation (sharing or collection). Categories with higher sensitivity weights contribute proportionally more to the overall score, ensuring that misalignment involving highly sensitive data e.g., financial or health information, has greater influence than misalignment involving low-risk data types.

To calculate the overall $\mathrm{SRS}$ for an app, we refer to $\mathrm{SRS}^{(\text{share})}_i$ as SRS-S and $\mathrm{SRS}^{(\text{collect})}_i$ as SRS-C. Thus, the overall scores is given as:
\begin{equation}
\label{eq:srs-overall}
\mathrm{SRS}^{\text{(overall)}}_i 
= \frac{\mathrm{SRS}^{(\text{share})}_i + \mathrm{SRS}^{(\text{collect})}_i}{2}
\end{equation}

Given that the $\mathrm{SRS}$ computes the expected misalignment intensity, it is important to emphasize sharing misalignment, as this implies that user data leave the device and therefore increases potential privacy risk. Accordingly, we define a revised overall score as follows:

\begin{equation}
\label{eq:srs-overall-w}
\mathrm{SRS}^{\text{(overall-w)}}_i 
= \alpha\,\mathrm{SRS}^{(\text{share})}_i 
 + (1 - \alpha)\,\mathrm{SRS}^{(\text{collect})}_i
\end{equation}

where $0 \le \alpha \le 1$ is a threshold parameter that controls the relative importance of alignment. 

\section{Cross-Layer Privacy Analysis - Quantitative Findings}
\label{results}
We now empirically evaluate the alignment between Google Play’s Data Safety labels and app privacy policies relying on the methodology and dataset described in 
Section~\ref{methodology}. This quantitative analysis is organized around the research questions presented in Section~\ref{rq}. 

\subsection{
RQ1. How consistent are privacy policies and Google Play’s Data Safety labels when reporting data collection and sharing?}
We first quantify consistency between privacy policy document (PPD) and the Data Safety label (DSL) across apps, data categories, and data practices (collection vs.\ sharing). We assess consistency using (i) overall Consistency and Misalignment Scores for agreements and discrepancies, and (ii) the Cohen’s kappa ($\kappa$) for chance-corrected agreement measure.

\begin{figure}[t]
\centering
\includegraphics[width=0.47\textwidth, height=5cm]{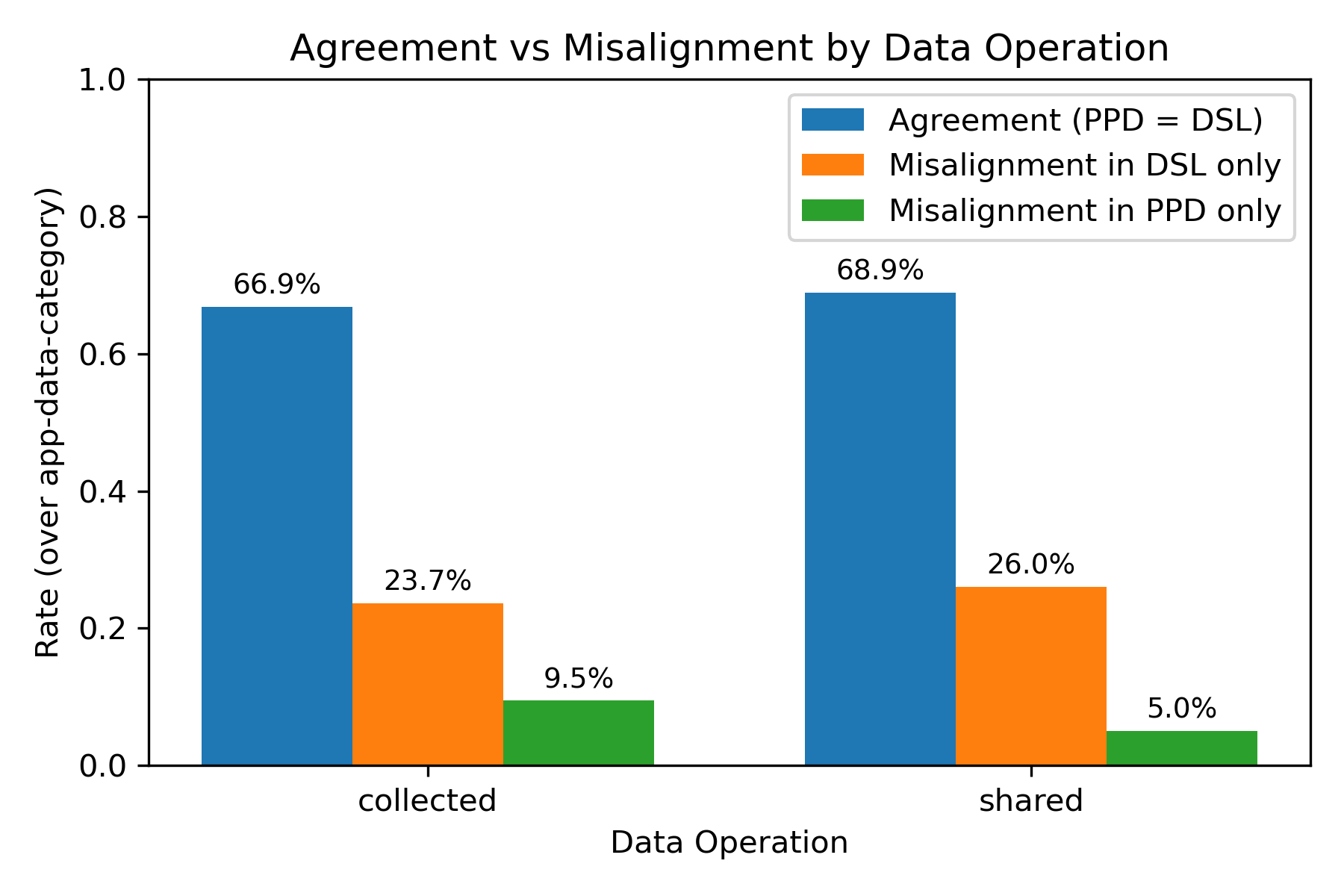}
\caption{Overall agreement vs.\ misalignment rates between PPD and DSL by operation (collection vs.\ sharing), aggregated over all apps and data categories.}
\Description{Overall agreement vs.\ misalignment rates between PPD and DSL by operation (collection vs.\ sharing), aggregated over all apps and data categories.}
\label{fig:misalignment_rates}
\end{figure}

\textbf{Overall Consistency and Misalignment:} Using the Consistency Score defined in Equation \ref{eq:consscore1}, we computed the PPD and DSL agreement across all apps and data categories. Our results show that the PPD and DSL agree at about 66.9\% for data collection disclosure and 68.9\% for data sharing disclosures. Since misalignment as defined in Equation \ref{eq:misscore} is the complement of the consistency indicator as shown in Equation \ref{eq:consscore2}, correspondingly, the overall misalignment rates (prevalence of disagreement across all app–category pairs) for our dataset are 33.1\% for data collection and 31.1\% for data sharing. However, among misalignment, discrepancies are more often attributed to the \emph{DSL only} than to the \emph{PPD only}. For collection, 23.7\% of disclosures are misaligned only in DSL versus 9.5\% only in PPD; for sharing, the corresponding rates are 26.0\% (DSL only) versus 5.0\% (PPD only). This asymmetry indicates that DSL-only discrepancies are systematically more common, especially for data sharing. This micro-level misalignment prevalence is shown in Figure \ref{fig:misalignment_rates}. At the macro-level, however, inconsistencies are more pervasive. Over 90.0\% of apps contain at least one inconsistency in DSL-reported data collection, and 92.6\% in DSL-reported data sharing. By comparison, inconsistencies in the PPD are substantially lower: 64.8\% of apps show at least one collection inconsistency, and only 35.4\% for sharing. This proportion reflects the significance of misalignment at the application level.

\begin{figure}[h]
\centering
\includegraphics[width=0.4\textwidth, height=5cm]{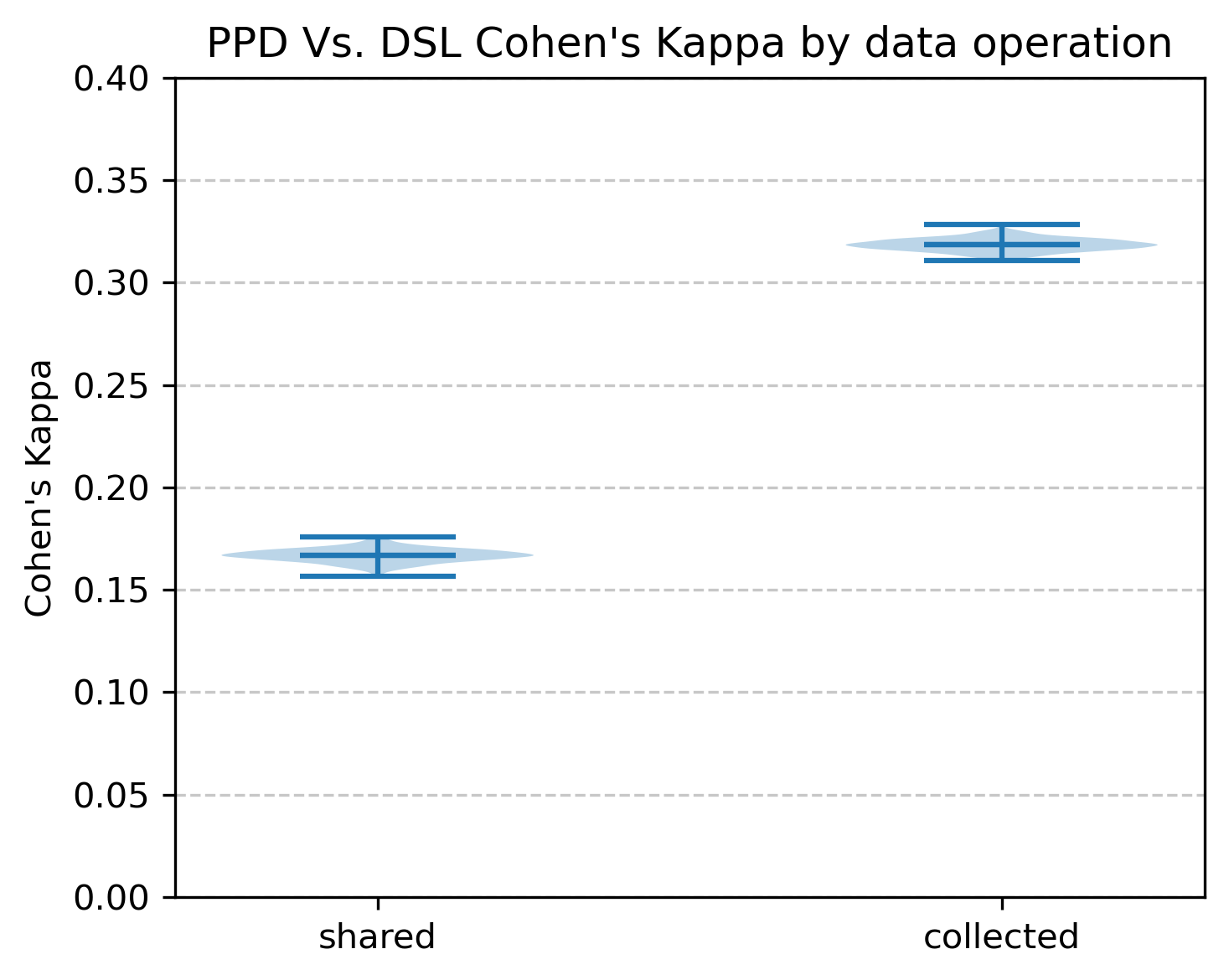}
\caption{PPD vs DSL kappa score by data operation scope}
\Description{PPD vs DSL Kappa Score by Data Operation Scope for sensitive data actions on data types.}
\label{fig:kappa}
\end{figure}

\textbf{Chance-corrected Consistency (Cohen’s $\kappa$).}
Although the consistency measures computed above showed that about 2 out of 3 app-category pairs agree at the micro-level, this measure did not account for chance-based agreement. In our dataset, we found that both sources often report "no data collected" or "no data shared" for many of the 14 data categories; using the simple rate-based measure, this will result in high prevalence of negative answers and would otherwise inflate agreement even if the two disclosures are poorly aligned. The $\kappa$ score, on the other hand, adjusts for this baseline and therefore provides a more reliable measure of true consistency across categories. Thus, to complement these rate-based prevalences, we compute Cohen's kappa ($\kappa$) to assess how consistently the PPD and DSL report collection and sharing, accounting for agreement that could occur by chance. Our result showed that the $\kappa$ scores average around 31\% for collection and 16\% for sharing, as shown in Figure \ref{fig:kappa}, indicating only weak agreement for collection and even weaker agreement for sharing. These low $\kappa$ values suggest limited consistency between PPD and DSL, especially in sharing.

\begin{tcolorbox}[left=2pt,right=2pt,top=2pt,bottom=2pt,colback=gray!5,colframe=gray!50,boxrule=0.5pt,title={\small RQ1 Key Findings}]
\small
Overall micro-level consistency rates are moderate (two-thirds), and chance-corrected consistencies are substantially lower (less than a third) for data sharing than for data collection. Additionally, macro-level inconsistencies are also widespread: almost all apps exhibit at least one disclosure misalignment. Taken together, these findings indicate that the two disclosure layers frequently present divergent representations of app data practices. Such divergence undermines the reliability of platform-level privacy disclosures and has significant implications for user data privacy, as users may base trust decisions on incomplete or inconsistent information.
\end{tcolorbox}

\subsection{RQ2. Which data categories are most frequently misaligned across the two disclosure layers?}
While RQ1 examines overall micro and macro app-level consistency/misalignment between PPD and DSL at the collection and sharing, RQ2 focuses on identifying the specific data categories with the highest misalignment prevalence. 

\begin{figure}[h]
\centering
\includegraphics[width=0.47\textwidth, height=7cm]{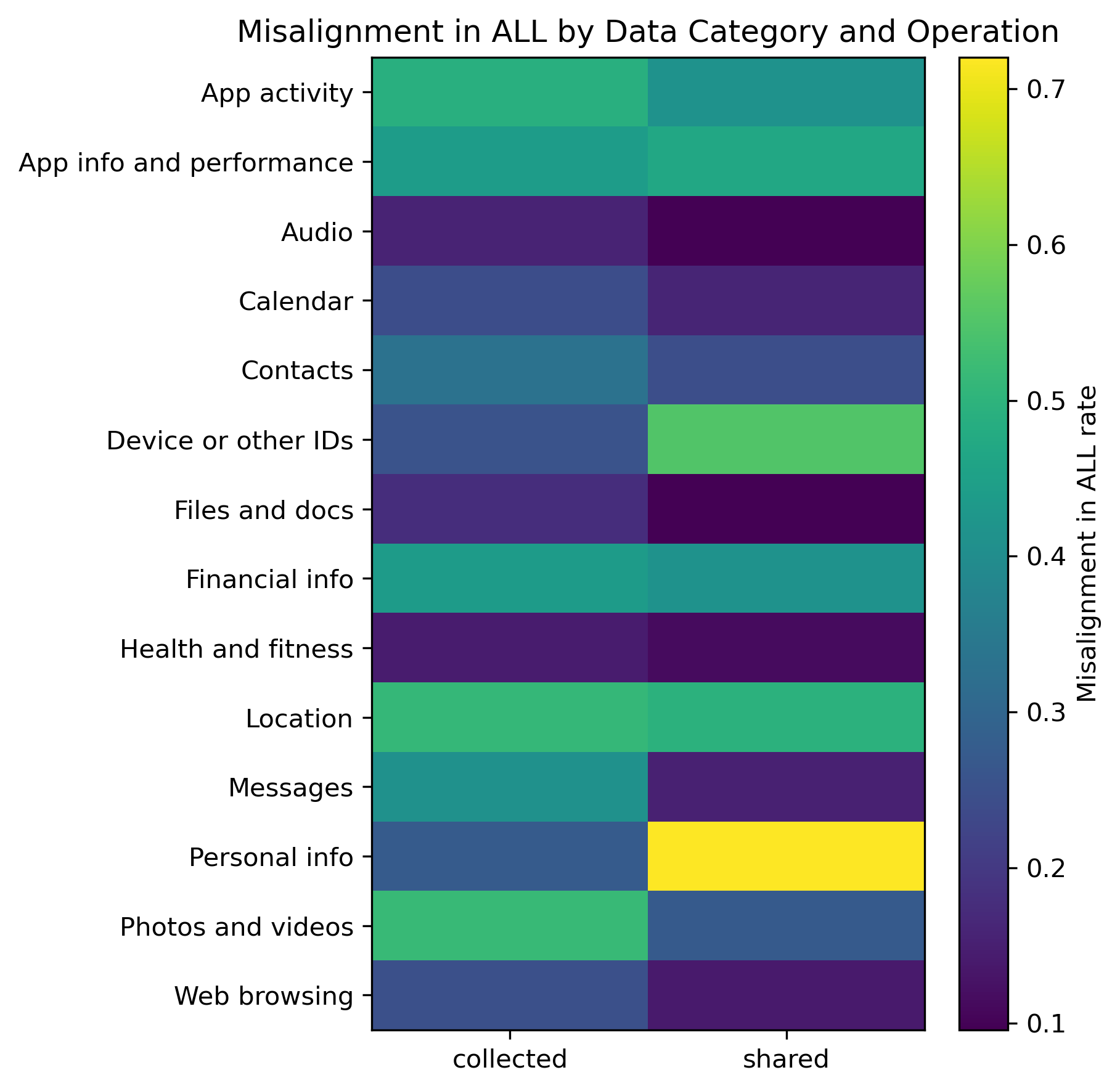}
\caption{Misalignment rates by data category and operation scope.}
\Description{A plot showing the ECDF of cosine similarity for sensitive data actions on data types.}
\label{fig:misalignment_heatmap}
\end{figure}

The heatmap in Figure~\ref{fig:misalignment_heatmap} shows category-level total misalignment rates for collection and sharing. Several categories exhibit strikingly high rates. For collection, the worst categories are Photos and videos (51.6\%), Location (51.1\%), App activity (49.0\%), App info and performance (44.0\%), Financial info (43.9\%), and Messages (41.1\%). For sharing, misalignment is even more pervasive for some categories: Personal info (72.0\%), Device or other IDs (55.1\%), Location (49.6\%), App info and performance (46.9\%), App activity (41.5\%), and Financial info (41.4\%). In contrast, categories such as Audio, Calendar and Health and fitness have noticeably lower misalignment rates in both operations.

\begin{figure}[h]
\centering
\includegraphics[width=0.47\textwidth, height=5.9cm]{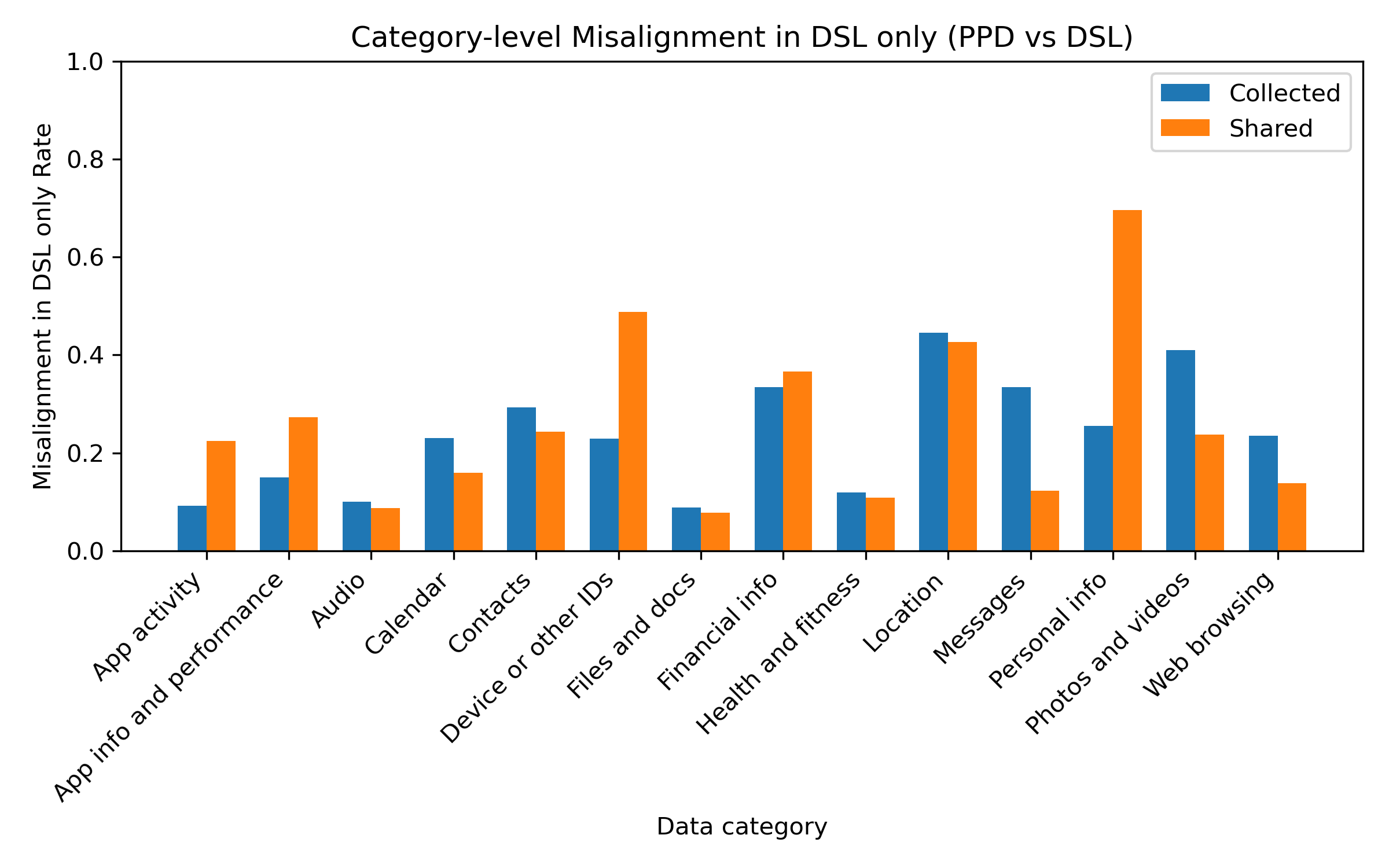}
\caption{Misalignment rates by data category in Data Safety labels only.}
\Description{A plot showing the ECDF of cosine similarity for sensitive data actions on data types.}
\label{fig:category_underreport}
\end{figure}

Additionally, to explore the specifics of the misalignment rate, we examine Misalignment in DSL (the Data Safety label omits collection/sharing described in the privacy policy) and Misalignment in PPD (the label claims collection/sharing that is not supported by the policy text) separately. Figure~\ref{fig:category_underreport} focuses on Misalignment in DSL, and the result shows that there is a significant misalignment in highly sensitive data categories. For example, 69.6\% of apps are misaligned in DSL for Personal info (shared). Similarly, for Location, misalignment in DSL is about 44.6\% for collection and 42.6\% for sharing; Device or other IDs (shared) show 48.8\%, and Photos and videos (collected) have a 41.0\% misalignment rate in DSL. Messages and Financial info have misalignment rates of about 33–37\% in DSL, depending on the operation. Taken together, these patterns indicate that for sensitive categories, misalignment largely reflects omissions in the Data Safety label relative to the policy. 

\begin{figure}[h]
\centering
\includegraphics[width=0.47\textwidth, height=5.9cm]{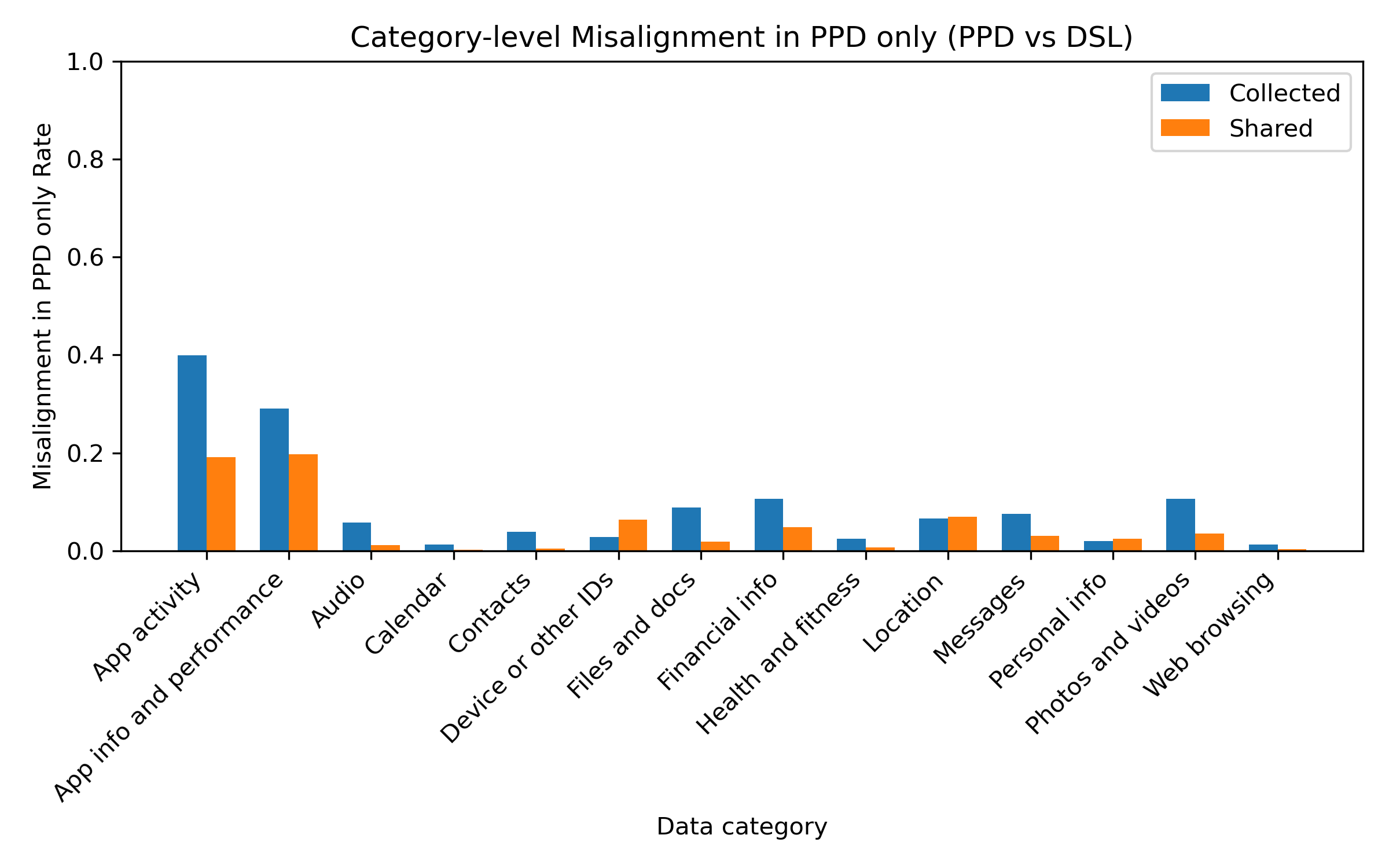}
\caption{Misalignment rates by data category in privacy policies only.}
\Description{A plot showing the ECDF of cosine similarity for sensitive data actions on data types.}
\label{fig:category_overreport}
\end{figure}
By contrast, Figure~\ref{fig:category_overreport} shows that misalignment rate in the privacy policy document is much rarer and concentrated in a small set of categories. The most notable cases are App activity (collected), where 39.9\% of apps are misaligned with privacy policies; App info and performance (collected) (29.0\%); and Photos and videos (collected) and Financial info (collected), which both exceed 10\%. For most other categories and for sharing in general, Misalignment in PPD stays below 7\%.

\textbf{Cosine Similarity.} 
To complement the category-level misalignment analysis, we examine whether PPD and DSL disclosures exhibit similar \emph{overall disclosure patterns} across the 14 data categories for both collection and sharing. We use cosine similarity to quantify \emph{distributional alignment}, measuring how closely the disclosure profiles in the privacy policy and the Data Safety label align for a given app and data operation.
Specifically, for each app and operation, we compute the cosine similarity between the vector of disclosed data-type terms (lexical) in the policy and the corresponding vector in the label. The resulting score ranges from 0 to 1, with higher values indicating greater overlap (agreement) in disclosed categories and lower values indicating misalignment. We summarize these similarity scores across apps using the empirical cumulative distribution function (ECDF), where each point on the curve represents the proportion of apps with similarity at or below that value.

\begin{figure}[h]
\centering
\includegraphics[width=0.4\textwidth,height=4.7cm]{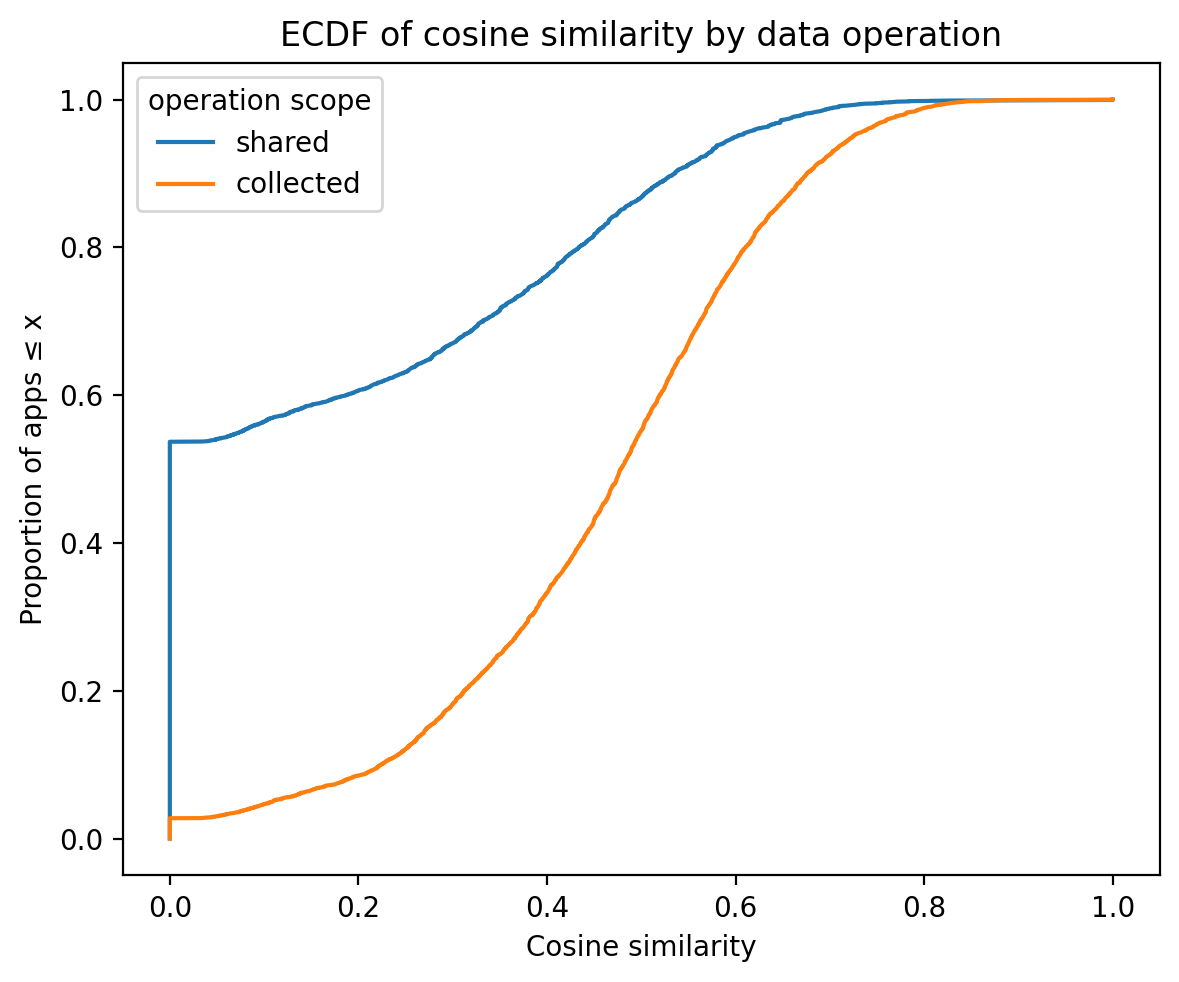}

\caption{ECDF of cosine similarity between privacy policies and Data Safety labels, stratified by data operation scope.}
\Description{A plot showing the ECDF of cosine similarity for data collection and data sharing across apps.}
\label{fig:cosine_similarity}
\end{figure}

As shown in Figure~\ref{fig:cosine_similarity}, the collection ECDF exhibits a sigmoidal (S-shaped) pattern, indicating that cosine similarity scores are concentrated around a central range rather than being uniformly distributed or strongly skewed. This suggests that, for most apps, Data Safety collection disclosures exhibit moderate structural alignment with the corresponding privacy policy descriptions. By contrast, the sharing curve rises sharply at very low similarity values, with approximately half of apps exhibiting cosine similarity near zero. This indicates a substantial subset of apps where reported sharing behavior in the Data Safety label has little to no lexical overlap with the privacy policy. This pattern is consistent with the frequent use of “No data shared with third parties” declarations in the Data Safety section, even when policies describe integrations with external partners. The median cosine similarity for collection is approximately 0.55, whereas the sharing distribution is heavily concentrated near the lower bound. Overall, developers appear substantially more consistent in describing data collection practices than data sharing practices. These aggregate trends are further illustrated by the case study in Appendix~\ref{openai}, where a “No data shared with third parties” coexists with extensive data sharing with external partners.

\begin{tcolorbox}[left=2pt,right=2pt,top=2pt,bottom=2pt,colback=gray!5,colframe=gray!50,boxrule=0.5pt,title={\small RQ2 Key Findings}]
\small
Overall, we find that high-sensitivity data categories exhibit the greatest misalignment between privacy policies and Data Safety labels. These discrepancies are driven primarily by under-reporting in the Data Safety label rather than over-disclosure relative to the policy. Consistent with this result, cosine similarity analysis also shows stronger distributional alignment for data collection than for data sharing, indicating that developers describe what apps collect more consistently than what they share. Because the most sensitive data categories are disproportionately affected—and sharing disclosures are particularly inconsistent—users may receive incomplete or misleading information about how and with whom their data are shared, undermining the reliability of data disclosures.\end{tcolorbox}

\subsection{RQ3. How are apps distributed across low, medium, and high Sensitivity Risk Score tiers, and what does this distribution reveal about the severity of privacy misalignment between privacy policies and Data Safety labels?}
RQ1 and RQ2 establish the prevalence of disclosure inconsistencies at both the app and data-category levels. We now extend this analysis by evaluating the privacy-sensitive impact of misalignment using the Sensitivity Risk Score (SRS) as defined in Equation~\ref{eq:srs-op}, which weights discrepancies according to the potential harm of the underlying data categories.

We assign sensitivity weights $w_c \in \{1,2,3\}$ corresponding to low, medium, and high sensitivity, to each of the 14 Google Play data categories. The categorization into three tiers is informed by prior literature characterizing the relative privacy risk associated with these data types. First, we treated \emph{Personal info} as highly sensitive, since prior disclosure-scoring work~\cite{aghasian2017scoring} assigns higher weights to core identifiers such as name, address, and contact details than to generic profile attributes. Next, we leverage the work of Sardana et al.~\cite{sardana2023ipds} and Chang et al.~\cite{chang2020framework}, which assign much higher sensitivity scores to permissions that access location, contact, financial, and health data than to diagnostics or performance metrics, to inform additional grouping decisions. Finally, drawing on the work of G{\'o}mez Ortega et al.~\cite{gomez2023sensitive}, which shows that voice interaction histories are often perceived as sensitive, we assigned audio a high sensitivity score in our scoring. Table~\ref{tab:sensitivity-weights} summarizes the final mapping used in our Sensitivity Risk Score, following the prior work characterizations.


\begin{table}[h]
\centering
\caption{Sensitivity weights assigned to Google Play data categories.}
\label{tab:sensitivity-weights}
\begin{tabular}{cl}
\hline
\textbf{Weight} & \textbf{Data categories} \\ \hline
3 &
Location, Personal info, Financial info, Audio, Contacts,\\
  & Health and fitness, Messages, Photos and videos, \\
  & Web browsing, Device or other IDs  \\[2pt]
2 &
Files and docs, App activity \\[2pt]
1 &
Calendar, App info and performance \\
\hline
\end{tabular}
\end{table}

\begin{figure}[h]
\centering
\includegraphics[width=0.5\textwidth, height=6.8cm]{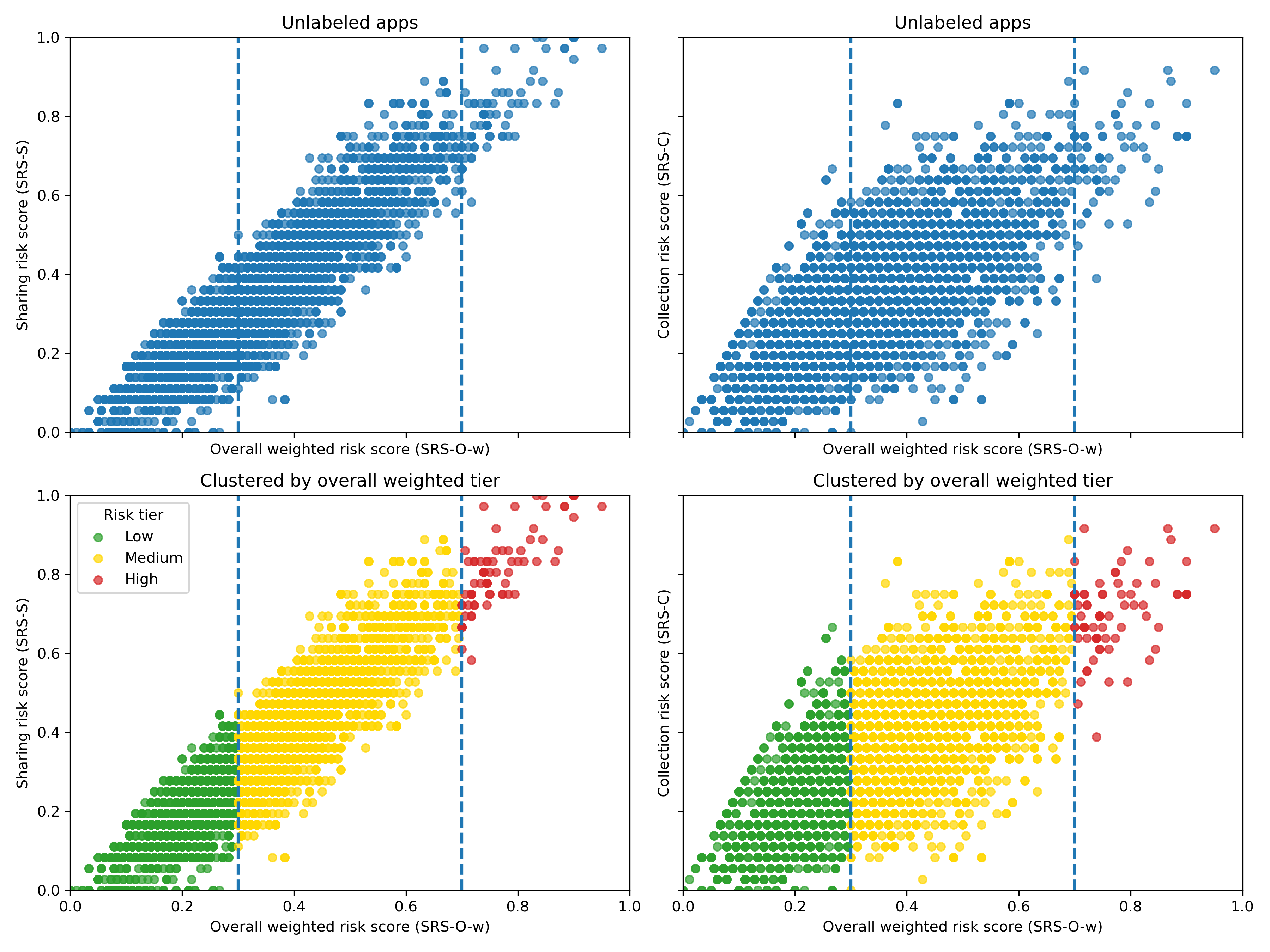}
\caption{Distribution of apps across risk tiers.}
\Description{A plot showing risk tiers for different apps.}
\label{fig:risktier}
\end{figure}

For the sensitivity risk scoring, we interpret each $\mathrm{SRS}^{(o)}_i$ variants: $SRS-C$ (collection) and $SRS-S$ (sharing), and $SRS-O-w$ (the overall weighted score -$\mathrm{SRS}^{\text{(overall-w)}}_i$) on a $0$--$1$ scale. To facilitate interpretation, we categorize apps into three risk tiers: low ($\text{score} < 0.30$), medium ($0.30 \le \text{score} < 0.70$), and high ($\text{score} \ge 0.70$). For the overall score defined in Equation~\ref{eq:srs-overall-w}, we set the parameter $\alpha$, which assigns moderately greater weight to sharing disclosures. As discussed in Section~\ref{sec:srs-definition}, this weighting reflects the greater privacy risk associated with transmitting data to third parties compared to collection alone. To evaluate the robustness of this choice, we conducted a sensitivity analysis using multiple $\alpha$ values, as reported in Appendix~\ref{alpha}. The results show that the overall risk stratification remains stable across the evaluated settings, with $\alpha = 0.6$ providing a balanced emphasis between collection and sharing disclosures. Using $\alpha = 0.6$, we computed the final SRS values for our dataset. A descriptive example of the SRS computation is provided in Appendix~\ref{ap:srs-eg}, and the resulting score distribution is illustrated in Figure~\ref{fig:risktier}.

The visualization in Figure~\ref{fig:risktier} shows the distribution of Sensitivity Risk Scores across all 6{,}051 apps by plotting the overall risk score ($SRS-O-w$) on the x-axis and the data collection and data sharing scores ($SRS-C$ and $SRS-S$) on the y-axes. Under $SRS-O-w$, apps are distributed between the low- and medium-risk tiers, with 49.61\% of apps in the low-risk tier and 49.10\% in the medium-risk tier, while only 1.29\% fall in the high-risk tier. Accordingly, the distribution is concentrated in the low- and medium-risk regions, with a comparatively small but visible tail extending into the high-risk region.
Apps located in the high-risk region generally exhibit elevated data sharing risk ($SRS-S$), and many also exhibit elevated data collection risk ($SRS-C$). This pattern suggests that high overall sensitivity-weighted risk typically reflects concentrated risk across multiple sensitive data practices, rather than being attributable to a single isolated factor. When comparing risk by data practice, high data-sharing risk is more prevalent than high data-collection risk. Specifically, 177 apps (2.93\%) fall in the high-risk tier under $SRS-S$, whereas 86 apps (1.42\%) fall in the high-risk tier under $SRS-C$. Comparing RQ3 with RQ2—which shows that high-sensitivity data categories are disproportionately affected by disclosure misalignment—the sensitivity-weighted SRS analysis reveals that substantial cumulative privacy risk is concentrated in a smaller subset of apps. In other words, while high-risk categories are more prone to misalignment, only a limited fraction of applications exhibit sustained or multi-category discrepancies severe enough to elevate their overall risk tier.

Importantly, the high-risk tier reflects not only the \emph{extent} of misalignment but also its \emph{sensitivity}. Under the weighting scheme in Table~\ref{tab:sensitivity-weights}, apps enter the high-risk region when discrepancies involve data categories with greater privacy harm potential. From a user-impact perspective, elevated risk in high-sensitivity categories—such as location, financial, health, communication, and identifier-related data—can increase exposure to profiling, identity linkage, behavioral tracking, discrimination, financial fraud, and disclosure of intimate personal attributes. Accordingly, the high-risk tier serves as a practical prioritization signal for auditing, developer review, and platform-level enforcement.\\

\begin{tcolorbox}[left=2pt,right=2pt,top=2pt,bottom=2pt,colback=gray!5,colframe=gray!50,boxrule=0.5pt,title={\small RQ3 Key Findings}]
\small
Overall, our results show that sensitivity-weighted scoring yields a stable and interpretable stratification of privacy risk across apps, while distinguishing how risk is distributed between data sharing and data collection. Notably, the distribution exhibits a more pronounced high-risk tail for data sharing than for data collection. These findings support $\mathrm{SRS}^{\text{(overall-w)}}_i$ as an effective measure of app-level privacy risk and demonstrate the value of tier-based stratification for prioritizing review of apps with risk concentrated in highly sensitive data categories.\end{tcolorbox}



\subsection{RQ4. Which app categories exhibit the highest misalignment risk, and how is risk distributed within those categories?}

\begin{figure}[h]
\centering
\includegraphics[width=0.47\textwidth, height=5.8cm]{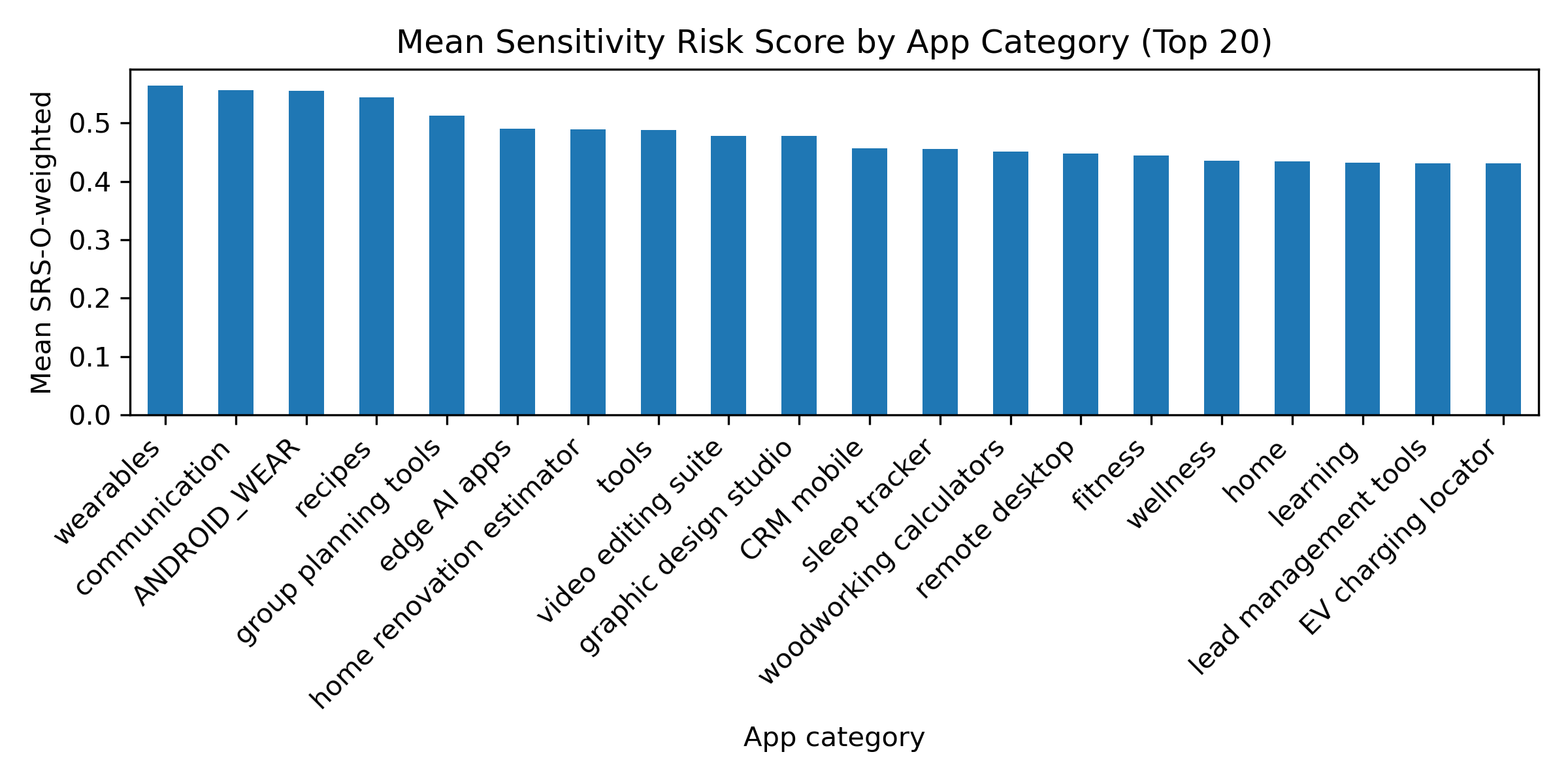}
\caption{Risk scores across app categories.}
\Description{A plot showing the ECDF of cosine similarity for sensitive data actions on data types.}
\label{fig:app_category}
\end{figure}

To investigate whether sensitivity risks are concentrated in particular kinds of apps, we analyze the SRS across Google Play app categories. For each category, we compute the mean overall weighted score $\mathrm{SRS}^{\text{(overall-w)}}_i$ and the distribution of apps across the low, medium, and high risk tiers.
Figure~\ref{fig:app_category} shows the top 20 categories ranked by mean overall weighted SRS. All of these categories have average scores between roughly 0.43 and 0.56, indicating that a typical app in these segments exhibits nontrivial misalignment between its privacy policy and data safety label. Categories related to communication and continuous data collection---such as \emph{wearables}, \emph{communication}, and \emph{ANDROID\_WEAR}---appear near the top of the ranking, as do several health- and lifestyle-related categories (\emph{sleep tracker}, \emph{fitness}, \emph{wellness}). We also observe
elevated scores for utility categories that can access devices or accounts remotely (e.g., \emph{remote desktop}, \emph{EV charging locator}), suggesting that apps with privileged or always-on access to users and devices are more prone to inconsistent disclosure.

\begin{figure}[h]
\centering
\includegraphics[width=0.47\textwidth, height=5.8cm]{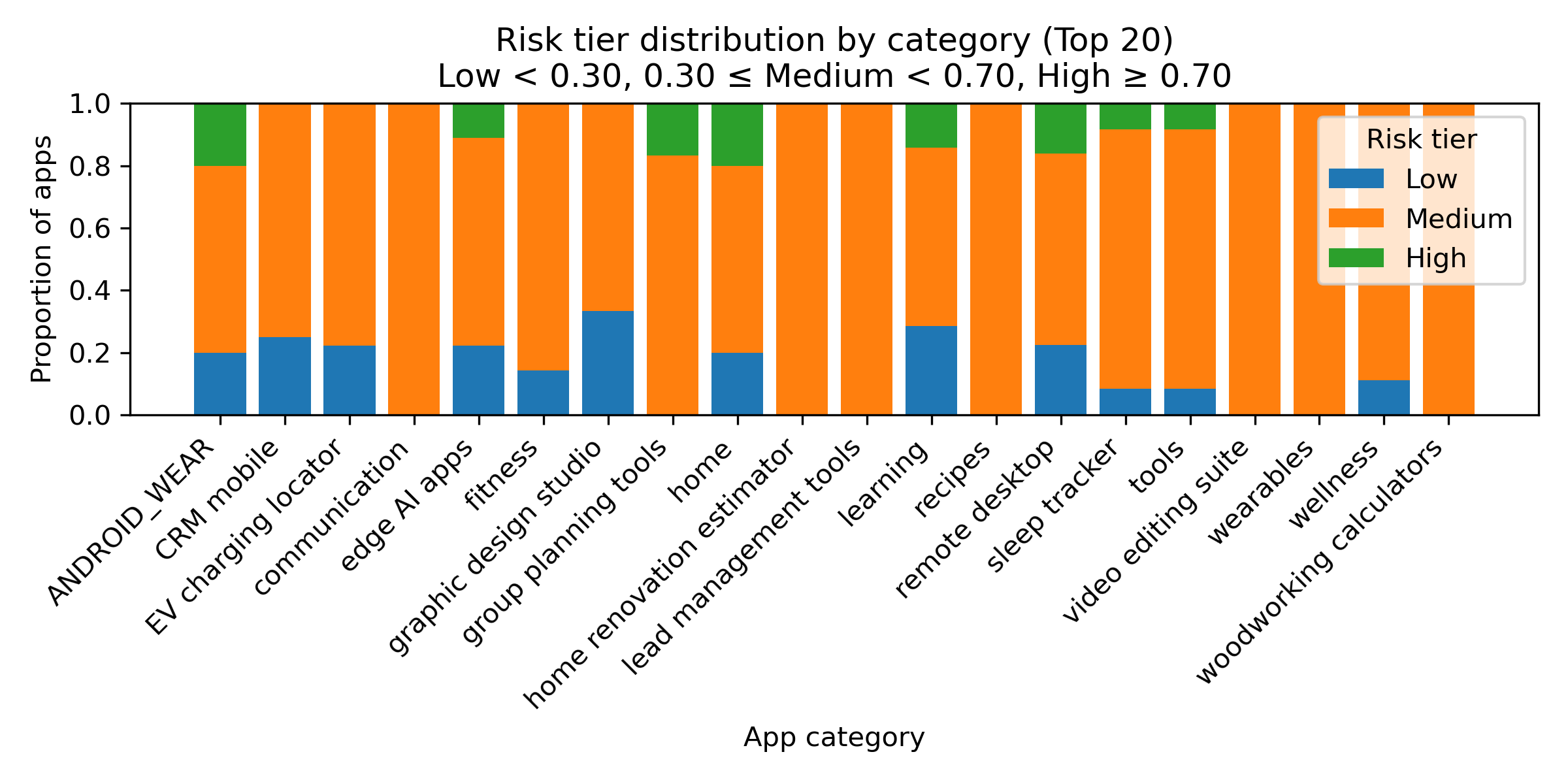}
\caption{Distribution of app categories across risk tiers.}
\Description{Distribution of App Categories Across Risk Tiers}
\label{fig:risktier_appcategory}
\end{figure}

Figure ~\ref{fig:risktier_appcategory} represents how apps are distributed across risk tiers within the top app categories shown in Figure ~\ref{fig:app_category}. In almost every category, the medium-risk tier dominates, confirming that moderate PPD--DSL misalignment is not an isolated occurrence. However, the share of high-risk apps is not uniform: categories such as \emph{ANDROID\_WEAR}, \emph{EV charging locator}, \emph{group planning tools}, and \emph{home renovation estimator} contain a visibly larger fraction of high-risk apps compared to others, while categories like \emph{wellness} or \emph{woodworking calculators} are almost entirely composed of low- and medium-risk apps. 
\begin{tcolorbox}[left=2pt,right=2pt,top=2pt,bottom=2pt,colback=gray!5,colframe=gray!50,boxrule=0.5pt,title={\small RQ4 Key Findings}]
\small
Using the sensitivity-weighted privacy risk score, we observe that app categories capable of persistent user monitoring, communication mediation, or device control exhibit significantly higher risk. This suggests that disclosure misalignment within these app types disproportionately involves high-sensitivity data categories, highlighting their potential privacy impact and underscoring the need for careful scrutiny of privacy disclosure claims in these domains.
\end{tcolorbox}

\noindent Additional quantitative evaluation of the correlation between privacy misalignment risk and app popularity is provided in Appendix \ref{ap:R5}.

\section{Qualitative Findings}
\label{discussion}

\subsection{Shared but Not Collected: Data Safety Labels}
In our subset of 3{,}111 apps without a privacy policy URL, we observed a recurring anomaly in which Google Play’s Data Safety label reports that an app shares user data while simultaneously declaring that it does not collect any data. Among these apps, 24.36\% indicate that user data are “shared” but that “no data” are collected. An example is \emph{GlobeOne: Get More from Globe}, whose store listing states that the app “may share” location, personal information, and device or other IDs with third parties, but also shows the summary “No data collected”~\cite{globeone2025}. According to Google’s own documentation, the Data Safety section is meant to disclose how apps collect, share, and protect user data, and developers are required to complete a form describing their practices, which is then shown on the store listing. Google explains that the label is developer-provided, and that “data collected” covers user data transmitted off the device, while “data shared” refers to data that is passed on to third parties \cite{google_datasafety_guide}, \cite{android_declare_data_use_2022}. Under these definitions, a pattern where multiple categories of personal and device data are marked as “shared” but overall “no data collected” appears logically inconsistent and suggests either misunderstanding of the form or misreporting of practices by the developer ~\cite{shah2022googledatasafety}. This type of labeling anomaly is important for our analysis because it illustrates how developer-provided disclosures can undermine the intended transparency of the Data Safety section and obscure the true extent of applications data-handling practices.

\subsection{The Service-Provider Loophole in Data Sharing Disclosures}
\label{service provider loophole}
Google Play’s Data Safety documentation allows developers to treat certain third parties as ``service providers'' rather than as recipients of ``shared'' data, as long as those entities process data solely on the developer’s behalf ~\cite{google_datasafety_guide}. For example, an analytics platform or cloud hosting provider can be classified as a service provider, and the associated data flows do not need to be disclosed as ``sharing'' in the Data Safety form. By contrast, Google explicitly states that if an SDK aggregates data across multiple customers to build advertising profiles, this is not service-provider activity and must be declared as data sharing in the label. In practice, this distinction can produce labels that say ``No data shared with third parties'' even when data is sent to large external companies. MX Player provides a concrete example: its privacy policy describes integrations with analytics, advertising, and SDK partners such as Google Analytics, Facebook Ads/AdMob, YouTube API Services, and Google APIs that collect information about installed apps and support interest-based advertising and profiling. If the developer classifies all of these partners as service providers, the Play Store listing can still show ``No data shared with third parties'' in the Data Safety summary, even though user data flows to those platforms~\cite{mxplayer_privacy,mxplayer_datasafety}. 

Under Google’s own guidance, however, any cross-app profiling or reuse of data across customers should be treated as sharing, implying that a ``No data shared'' label would be misleading in such a scenario. This pattern is consistent with our large-scale analysis in Section~\ref{results}, where we find that sharing disclosures are much less aligned with privacy policies than collection disclosures. This highlights a gap in how Google supports both users and developers. For users, the current label design does not reveal which external companies receive their data or whether those companies reuse it across apps, making it difficult to understand the true extent of data disclosure. For developers, the burden of correctly interpreting nuanced concepts such as ``service provider'' and ``cross-app profiling'' is placed entirely on them, while Google simultaneously emphasizes that it does not systematically verify the accuracy of declarations ~\cite{google_datasafety_guide}. We also observed data types not captured by the Data Safety categories; examples are provided in Appendix \ref{srs_robust}.

\subsection{Recommendations}
Our findings underscore the importance of strengthening Data Safety disclosure mechanisms on platforms such as Google Play. Ensuring closer alignment with privacy policies—particularly for sensitive data categories—is critical to reducing privacy-sensitive disclosure risk and providing users with an accurate understanding of data practices.

\subsubsection{Implement Automated Consistency and Sensitivity Scoring at Submission Time}
Beyond strengthening disclosure design, we recommend that Google integrate an automated cross-layer consistency verification mechanism at the time of app submission.
Specifically, the platform could:
(i) Automatically compare a developer’s privacy policy against the submitted Data Safety label using a structured category mapping similar to the methodology presented in this study. (ii) Compute and internally validate a Consistency Score and a Sensitivity Risk Score (SRS) that quantify the degree and risk-weighted impact of disclosure misalignment. (iii) Surface these scores transparently to users as part of the app listing interface.
Displaying a consistency indicator would provide users with an interpretable signal of disclosure reliability. Over time, greater visibility of low-consistency or high-risk scores may incentivize developers to improve alignment between privacy policies and Data Safety labels, creating market-driven pressure toward more accurate and trustworthy disclosures.
\subsubsection{Strengthen the Design and Transparency of the Data Safety Framework}
The results from this study indicate that disclosure misalignment—particularly involving sensitive data categories—often stems from limited transparency around third-party SDK practices and ambiguous service-provider classifications. To address this, we recommend that Google strengthen the design and enforcement of the Data Safety framework through the following measures: (i)
Require explicit identification of analytics, advertising, and other high-impact SDK partners within the Data Safety form, even when developers classify them as service providers. This should include disclosure of whether such partners reuse data across multiple applications. (ii) Surface SDK-level data accesses directly within the Data Safety interface, linking specific data categories (e.g., location, identifiers, financial data) to the third parties that receive them. This prevents developers from generically labeling all partners as service providers without concise differentiation. (iii) Establish standardized incident reporting mechanisms that map exposed data fields back to the exact categories and collection/sharing operations presented in the Data Safety label, improving accountability and post-incident transparency.


\subsection{Limitations and Future Work}
While our framework enables large-scale analysis of cross-layer privacy disclosure consistency, several limitations should be considered when interpreting the results. In this context, terms such as “under-reporting” refer to relative disclosure differences between the two artifacts; for example, when a data category is disclosed in the privacy policy but omitted from the corresponding Data Safety label. Such discrepancies do not necessarily establish that the label is factually incorrect or that the application actually collects or shares the data in practice. Rather, they indicate divergence between two developer-provided representations of data practices. Our privacy-policy analysis relies on an LLM-based extraction pipeline. We validated the pipeline on a manually labeled subset and observed stable performance across two high-performing LLM backends as shown in Section~\ref{backend validation}. Nevertheless, extraction quality may vary across models, policy-writing styles, ambiguous legal language, or future model versions. In addition, our crawler is currently restricted to English-language privacy policies, which may limit generalizability to non-English, multilingual, or region-specific app ecosystems where disclosures may be expressed differently. However, because our framework relies on general-purpose LLM extraction, extending the pipeline to multilingual settings is a feasible direction for future work. Finally, our framework intentionally represents disclosures using category-level binary indicators over the 14 Google Play data categories to support consistent, scalable, and reproducible comparison across thousands of apps. While appropriate for measuring broad cross-layer disclosure alignment, this abstraction does not capture all contextual nuances present in privacy disclosures, such as processing purpose, optional versus required collection, retention conditions, user controls, or fine-grained distinctions within broader data categories. Consequently, some partial or conditional disclosures may not be fully represented in the binary encoding.

As future work, we plan to extend our framework to a three-way analysis that triangulates static code analysis, privacy policies, and Data Safety labels. Such an approach could help distinguish benign disclosure inconsistencies from systematic under-disclosure and support more targeted privacy auditing and platform enforcement. Another important direction is longitudinal measurement. By periodically re-crawling the same applications, future work could examine how Data Safety labels and privacy policies evolve over time in response to policy changes, enforcement actions, platform redesigns, or public privacy incidents. Finally, future work could extend this cross-layer analysis to the Apple ecosystem by comparing Apple Privacy Nutrition Labels with privacy policies and, where possible, analyzing cross-platform versions of the same applications.

\section{Related Work}
\label{RW}
In this section, we review the related literature in automated privacy policy analysis and measurement studies evaluating the accuracy and usability of app-store privacy labels. 
\subsection{Privacy Policy Analysis}
Early efforts to automate privacy policy analysis used NLP and conventional ML to reduce the burden of reading long policies and to support scalable auditing.  The development of annotated corpora proved fundamental to this research direction, with Wilson et al.~\cite{wilson2016creation} introduced the OPP-115 dataset of 115 policies labeled by law students, which became a widely used benchmark for training and evaluating policy understanding systems \cite{wilson2018analyzing}. Building on such resources, several systems extracted structured representations of data practices. PolicyLint~\cite{andow2019policylint} used rule-based NLP to detect inconsistencies and potential contradictions in policy statements, while PrivOnto~\cite{oltramari2018privonto} mapped expert annotations into an ontology to support query knowledge bases of collection, sharing, and purpose. Yu et al.~\cite{yu2016can} similarly leveraged syntactic parsing and pattern matching to surface critical sentences about data handling. In parallel, statistical classifiers framed policy analysis as supervised text classification or completeness assessment. Costante et al.~\cite{costante2012machine} and Guntamukkala et al.~\cite{guntamukkala2015machine} applied multiple ML models to categorize policy content and assess coverage relative to regulatory-inspired categories. Later systems improved scalability and output structure such as Polisis~\cite{harkous2018polisis} used hierarchical neural models to map policy text into structured, human-readable labels, and PrivacyCheck~\cite{zaeem2018privacycheck} used TF--IDF features with gradient-boosted classifiers to generate aggregate privacy scores. Other work targeted specific policy elements, such as opt-out and choice language~\cite{sathyendra2017identifying}. In general, these prior methods rely on task-specific schemas, extensive manual annotation, and brittle rule-based extraction pipelines. 

These limitations have motivated a new line of research leveraging LLM-assisted policy analysis frameworks to automatically extract data collection and sharing statements from privacy policies into normalized schemas suitable for large-scale measurement. The works of \cite{zaeem2018privacycheck,goknil2024privacy,wu2025clear,sowe2024design,qiu2024embodied,palka2025make} have shown strong LLM performance in policy summarization. Additionally, \cite{mori2022analysis,chadwick2024assessments,freiberger2025you,aydin2024assessing,amaral2021ai,torre2020ai} proposed the use of LLMs for regulatory-focused policy assessment and compliance checking from privacy policy. Woodring et al.'s \textit{Privacify} \cite{woodring2024enhancing} leverages LLMs to summarize policies and support regulation-oriented analysis (e.g., GDPR, HIPAA, COPPA, FERPA) by generating structured outputs and legal rationales. Rodriguez et al. \cite{rodriguez2024large} further demonstrate that general-purpose LLMs such as ChatGPT and Llama~2 can compete with and often outperform state-of-the-art traditional methods for extracting privacy practice disclosures, suggesting a paradigm shift towards efficient and accessible analysis. Beyond extraction and regulatory mapping, LLMs can also support user-centered understanding by mitigating information overload; for example, Mori et al. \cite{mori2025evaluating} use LLMs to evaluate and improve policy understandability, showing that these models can identify policy descriptions that users are likely to misunderstand. Tang et al. \cite{tang2023policygpt} further introduced \textit{PolicyGPT}, an LLM-based framework for classifying policies with respect to GDPR requirements. Our work builds on this foundation of LLM-assisted policy analysis to investigate cross-layer consistency between privacy policies and platform-level Data Safety disclosures. 

In contrast, we move beyond structured extraction by extending the LLM-assisted policy analysis framework proposed by \cite{woodring2024enhancing} to support systematic cross-layer disclosure alignment measurement. In addition, we introduce a sensitivity-weighted scoring framework that quantifies the privacy risk associated with disclosure misalignment.

\subsection{Data Safety Labels Evaluation}
\label{sec:rw_dss_reliability}
With the mandatory introduction of short-form privacy disclosures, including Apple’s Privacy Nutrition Labels and Google Play’s Data Safety section, recent research has examined whether these labels serve as reliable transparency mechanisms by evaluating their accuracy and how effectively they support users’ privacy decisions. On iOS, Ali et al.~\cite{ali2024honesty} used BERT-based language models to compare Apple privacy labels against privacy policies and found widespread mismatches in which policies often described broader data collection than what labels reported. Using behavioral evidence instead of policies, Kollnig et al.~\cite{kollnig2022goodbye} evaluated Privacy Nutrition Labels using static analysis and network-traffic monitoring, showing that apps labeled as not collecting data can still include tracker libraries and exhibit tracking-related behavior. Cross-ecosystem comparisons further reveal inconsistency even for the same app: Rodriguez et al.~\cite{rodriguez2023comparing} compared disclosed data collection across five data categories and found that iOS and Google Play labels frequently disagree. Focusing specifically on Android, Khandelwal et al.~\cite{khandelwal2024unpacking} conducted the first comprehensive measurement of Google Play’s Data Safety Section, identifying substantial internal inconsistencies and evidence of both under- and over-reporting. To explain how such issues arise, their developer study highlights practical reporting challenges such as ambiguous terminology, difficulty accounting for SDK-driven data practices, and the burden of keeping disclosures updated, which aligns with Li et al.~\cite{li2022understanding}, who observed and interviewed developers completing Apple’s labels and found that misunderstandings of label terms and uncertainty about what to disclose are common sources of inaccuracies. Beyond correctness, user studies suggest that labels often fail to meet user information needs. Zhang et al.~\cite{zhang2023privacy} showed that current labels answer only a minority of users’ real privacy questions, particularly omitting details such as who receives shared data and who can access it, while Zhang et al.~\cite{zhang2022usable} found that users struggle with label structure and technical terminology and that labels have limited impact on users’ ability to manage privacy. In contrast to prior work, our study introduces a cross-layer privacy disclosure analysis pipeline that normalizes privacy-policy and Data Safety disclosures into a unified schema, enables large-scale alignment analysis of data collection and sharing practices, and ranks disclosure misalignment using a sensitivity-weighted risk score.


\subsection{Cross-Layer Privacy Disclosure Studies}
Recent studies, including Alomar et al.~\cite{alomar2025effect} and Jain et al.~\cite{jain2023atlas}, have examined inconsistencies between platform privacy labels and privacy policies in both Android and iOS ecosystems. 
 ATLAS~\cite{jain2023atlas} studies discrepancies between Apple Privacy Labels and privacy policies at scale in the iOS ecosystem using an ensemble classifier for privacy-label prediction, while Alomar et al.~\cite{alomar2025effect} analyze behavioral compliance and inaccurate disclosures in child-directed Android apps using dynamic traffic analysis and SDK inspection. Our work differs in both analytical scope and methodological focus. We focus on cross-layer consistency between privacy policies and Google Play Data Safety Labels across 6,051 Android apps using a unified 14-category Google Play schema. Methodologically, we leverage an LLM-based extraction framework to identify explicit collection and sharing disclosures directly from privacy policies and compare them against developer-declared DSL disclosures.

Beyond identifying disclosure discrepancies, our study introduces additional analytical dimensions for characterizing disclosure misalignment. First, we explicitly distinguish data collection and data sharing as separate disclosure operations throughout the analysis. This enables operation-specific consistency measurements and reveals a strong asymmetry between collection and sharing disclosures, with sharing disclosures exhibiting substantially lower consistency. Second, our framework decomposes discrepancies into DSL-only and PPD-only mismatches, enabling characterization of systematic omission patterns within Google Play Data Safety Labels, particularly for sharing-related disclosures. Third, we introduce a Sensitivity Risk Score that weights discrepancies according to the sensitivity of the underlying data categories, allowing misalignment severity to be analyzed beyond raw discrepancy frequency alone. Finally, we incorporate chance-corrected agreement analysis using Cohen’s ($\kappa$) and structural disclosure alignment using cosine similarity to characterize agreement beyond raw match rates. 

Collectively, these analyses provide a more fine-grained and risk-oriented characterization of cross-layer disclosure misalignment across the Google Play ecosystem, revealing that sharing disclosures are systematically less aligned than collection disclosures and that elevated disclosure risk disproportionately concentrates in high-sensitivity data categories.

\section{Conclusion}
\label{conclusion}

This study examined how Android developers’ privacy policies align with their Google Play Data Safety labels regarding user data-handling disclosures. Using a unified schema across 14 data categories and a sensitive privacy risk score, our analysis of over 6000 apps reveals widespread inconsistencies between privacy policies and data safety labels, particularly regarding data sharing. Under-reporting in the data safety label is much more common, and mismatches are concentrated in sensitive categories such as personal information, location, financial data, device identifiers, and messages. Our risk analysis further indicates that around half of the apps fall into a medium-risk tier, and a fraction, but an important subset, appear high risk. Overall, current developer self-reported labels provide only a partial and sometimes misleading picture of user data practices, underscoring the need for stronger platform checks and more transparent, verifiable disclosure mechanisms.


\begin{acks}
This research received no specific grant from any funding agency in the public, commercial, or not-for-profit sectors. The authors used generative AI-based tools to revise the text, correct typographical errors, and grammatical errors.
\end{acks}
\balance
\bibliographystyle{ACM-Reference-Format}
\bibliography{ref}

\appendix
\clearpage
\appendix

\section{Data Safety Label Example}
\label{DSLEx}
\begin{figure}[h]
\centering
\setlength{\fboxsep}{0pt}      
\setlength{\fboxrule}{0.2pt}   
\fbox{%
  \includegraphics[width=0.25\textwidth,height=6cm]{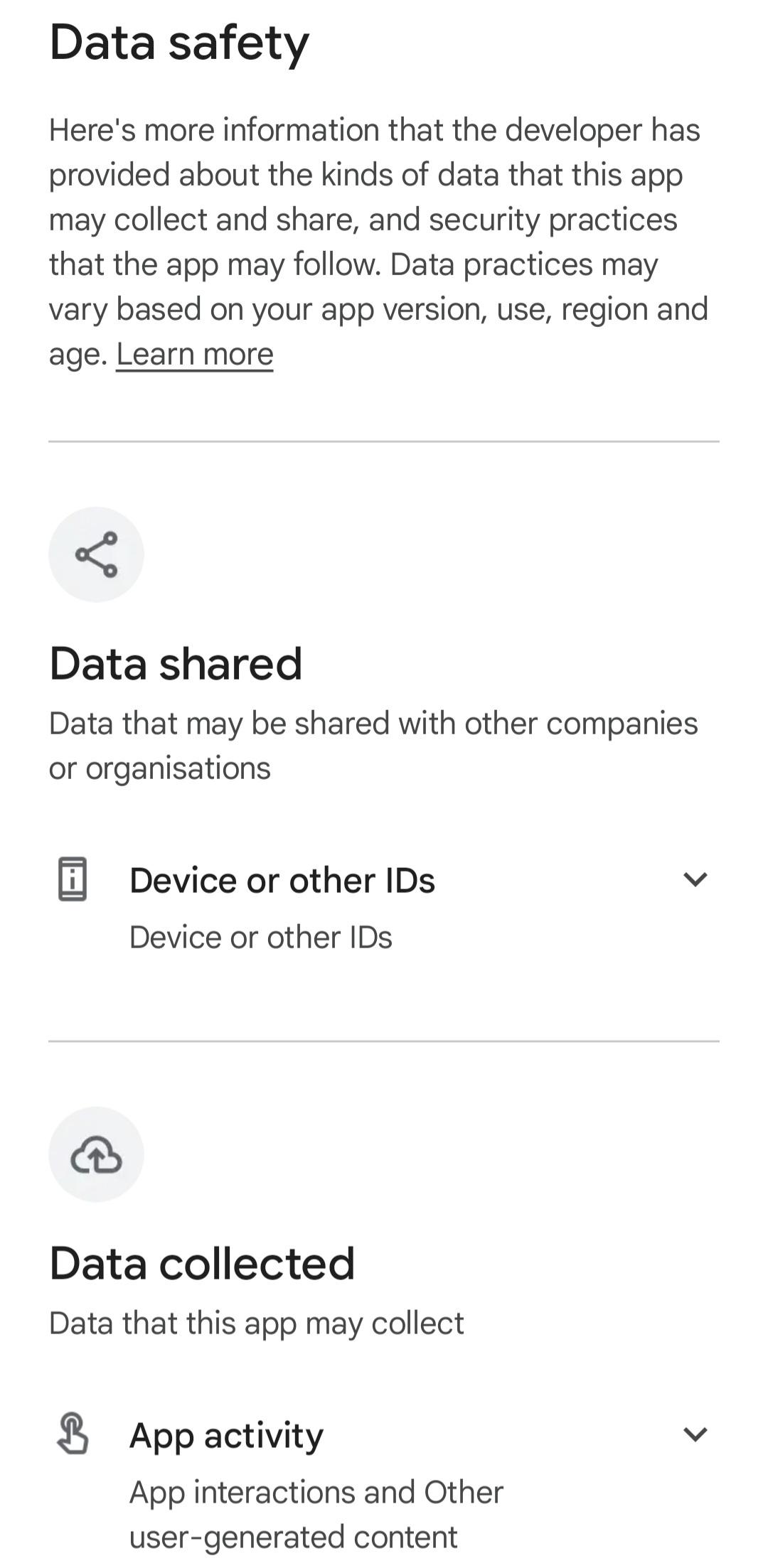}%
}
\caption{An Example of Data Safety Label}
\Description{data safety label example}
\label{fig:dsl}
\end{figure}

\section{Data Safety Label Data Structure Example}
\label{dslex}

\begin{lstlisting}[style=jsonstyle,
caption={Data Safety label for the \texttt{Bluesky} app},
label={lst:bluesky-dsl}]
{
  "Bluesky": {
    "Data shared": {
      "App info and performance": "Crash logs, Diagnostics, and Other app performance data",
      "Device or other IDs": "Device or other IDs",
      "Personal info": "User IDs",
      "App activity": "App interactions and Other user-generated content"
    },
    "Data collected": {
      "App info and performance": "Crash logs, Diagnostics, and Other app performance data",
      "Device or other IDs": "Device or other IDs",
      "Photos and videos": "Photos",
      "Personal info": "Email address and User IDs",
      "App activity": "App interactions and Other user-generated content"
    }
  }
}
\end{lstlisting}

\section{Privacy Policy Data Structure Example}
\label{ppdEx}

\begin{lstlisting}[style=jsonstyle,
caption={Data types collected and shared with third parties according to the Bluesky privacy policy},
label={lst:bluesky-policy-datatypes}]
{
  "Bluesky": {
    "collected": [
      "email address",
      "phone number",
      "images you associate with your profile",
      "birth date",
      "username",
      "name",
      "posts",
      "comments",
      "direct messages",
      "any personal information you provide with your application",
      "payment information",
      "card information",
      "Internet protocol (IP) address",
      "user settings",
      "cookie identifiers",
      "mobile carrier",
      "other unique identifiers",
      "browser or device information",
      "Internet service provider (ISP)",
      "the posts you view on Bluesky",
      "the links you click within Bluesky",
      "the frequency and duration of your activities",
      "password",
      "birthday",
      "telephone number",
      "government identifiers",
      "Age",
      "address",
      "passport",
      "account information",
      "driver's licence number",
      "passport number",
      "partial debit card information",
      "partial credit card information",
      "bank account information",
      "other payment or financial information",
      "date of birth",
      "gender",
      "purchase history",
      "device information",
      "log data",
      "pictures and videos (content) you upload",
      "role with the business or organisation",
      "communication preferences",
      "chat history",
      "messages you send us",
      "generated content"
    ],
    "shared": [
      "email address",
      "phone number",
      "profile image",
      "birth date",
      "username",
      "name",
      "posts",
      "comments",
      "direct messages",
      "payment information",
      "card information",
      "Internet protocol (IP) address",
      "user settings",
      "cookie identifiers",
      "mobile carrier",
      "other unique identifiers",
      "browser or device information",
      "Internet service provider (ISP)",
      "the posts you view on Bluesky",
      "the links you click within Bluesky",
      "the frequency and duration of your activities",
      "password",
      "birthday",
      "telephone number",
      "government identifiers",
      "age",
      "address",
      "passport",
      "account information",
      "driver's licence number",
      "passport number",
      "partial debit card information",
      "partial credit card information",
      "bank account information",
      "other payment or financial information",
      "date of birth",
      "gender",
      "purchase history",
      "device information",
      "log data",
      "pictures and videos (content) you upload",
      "role with the business or organisation",
      "communication preferences",
      "chat history",
      "messages you send us",
      "generated content"
    ]
  }
}
\end{lstlisting}

\section{SRS Calculation in Practice}\label{ap:srs-eg}
To illustrate how the SRS is calculated in practice, we provide an example for a single app under the proposed weighting scheme. Consider an app $x$ with three data categories $C_x=\{i,j,k\}$.
Assume sensitivity weights are $w_i=3$ (high), $w_j=2$ (medium), and $w_k=1$ (low).
For the \textit{collection} operation, suppose the policy and label agree on categories $j$ and $k$ but disagree on $i$.
Then $\mathrm{Cons}^{(\text{collect})}_{x,i}=0$ and $\mathrm{Cons}^{(\text{collect})}_{x,j}=\mathrm{Cons}^{(\text{collect})}_{x,k}=1$.
By Eq.~\eqref{eq:srs-op},
\[
\mathrm{SRS}^{(\text{collect})}_x
=\frac{3(1-0)+2(1-1)+1(1-1)}{3+2+1}
=\frac{3}{6}=0.50.
\]
For \textit{sharing}, assume the app disagrees on $i$ and $j$ but agrees on $k$,
so $\mathrm{Cons}^{(\text{share})}_{x,i}=0$, $\mathrm{Cons}^{(\text{share})}_{x,j}=0$, and $\mathrm{Cons}^{(\text{share})}_{x,k}=1$.
Then
\[
\mathrm{SRS}^{(\text{share})}_x
=\frac{3(1-0)+2(1-0)+1(1-1)}{6}
=\frac{5}{6}\approx 0.83.
\]
The unweighted overall score (Eq.~\eqref{eq:srs-overall}) is
$\mathrm{SRS}^{(\text{overall})}_x=(0.50+0.83)/2\approx 0.67$.
Using the weighted overall score (Eq.~\eqref{eq:srs-overall-w}) with $\alpha=0.6$,
\[
\mathrm{SRS}^{(\text{overall-w})}_x
=0.6(0.83)+0.4(0.50)\approx 0.70,
\]
placing app $x$ at the boundary of the \textit{high-risk} tier under our thresholds.

\section{Additional Quantitative Analysis}\label{ap:R5}
RQ5. How does privacy misalignment risk relate to app popularity (user ratings and installs)?

\begin{figure}[h]
\centering
\includegraphics[width=0.5\textwidth, height=5cm]{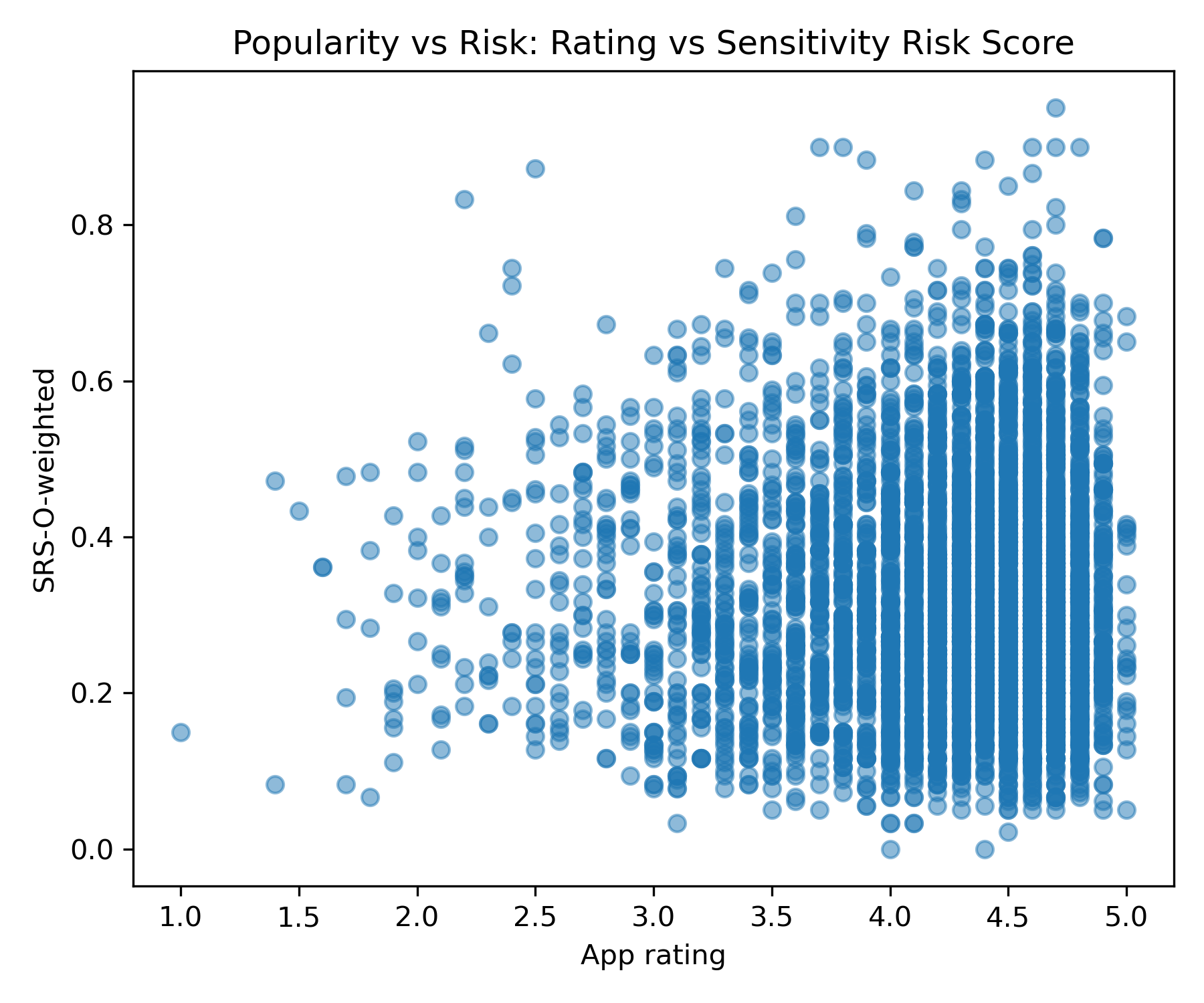}
\caption{Rating Vs Risk Score}
\Description{A plot showing the rating vs risk score}
\label{fig:rating}
\end{figure}

To examine whether privacy-sensitive disclosure risk is associated with app popularity, we correlate the overall weighted Sensitivity Risk Score (SRS) with app rating and download count. Figure~\ref{fig:rating} plots the overall weighted score $\mathrm{SRS}^{\text{(overall-w)}}_i$ against the app’s average Google Play rating. The points form a broad cloud rather than a clear trend: across the common rating range of 3.0–4.8, we observe a full spectrum of risk scores, from near 0 to above 0.8. In particular, many highly rated apps (4+ stars) exhibit elevated privacy-sensitive risk scores, indicating substantial misalignment in the disclosure of sensitive data categories. Visually, there is no clear evidence that higher user ratings are associated with lower SRS values. This suggests that privacy-sensitive disclosure risk is either not readily observable to users or does not systematically influence their rating behavior.

\begin{figure}[h]
\centering
\includegraphics[width=0.5\textwidth, height=6cm]{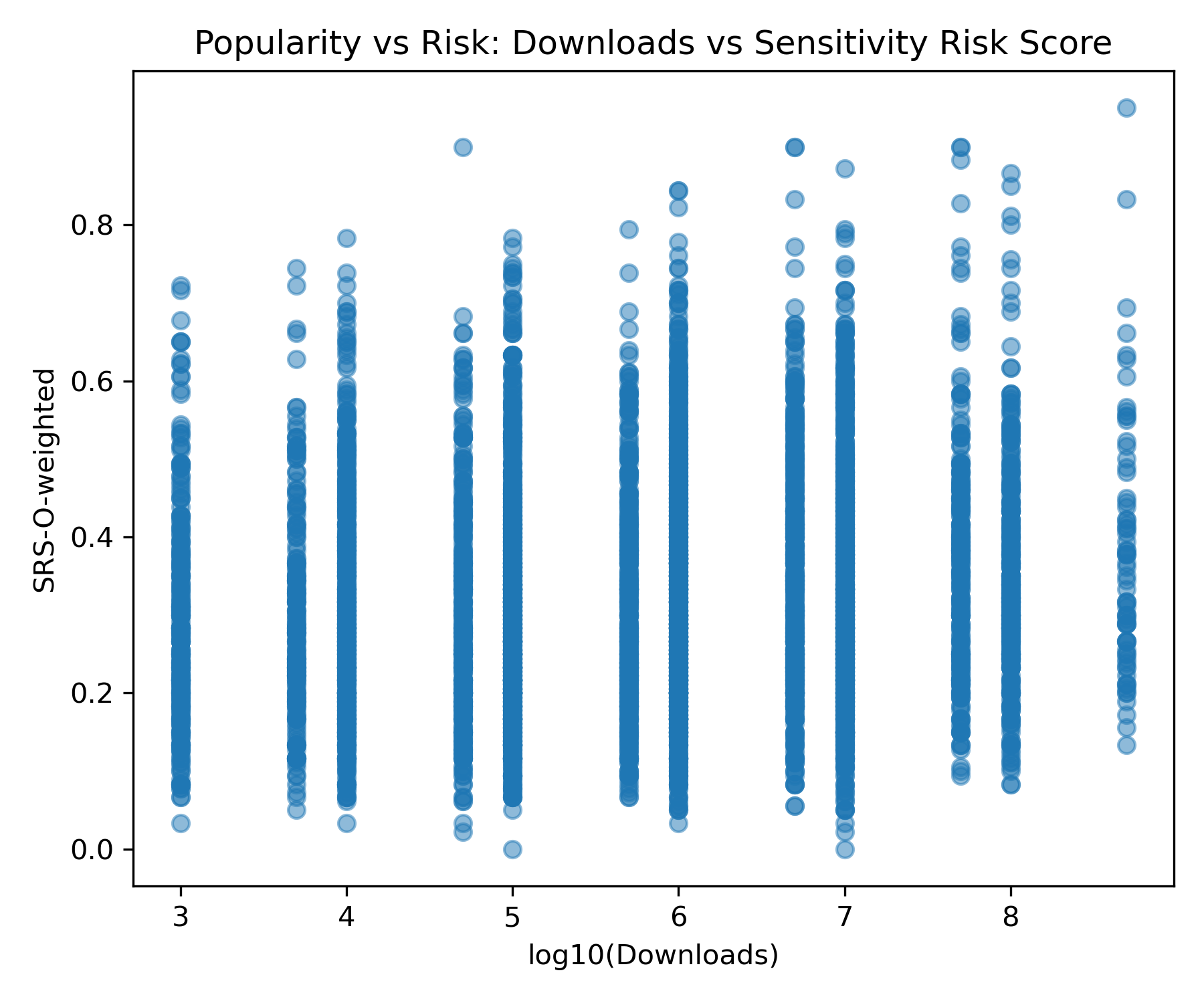}
\caption{Download Vs Risk Score}
\Description{A plot showing the ECDF of cosine similarity for sensitive data actions on data types.}
\label{fig:download}
\end{figure}
Figure~\ref{fig:download} shows a similar analysis for app downloads, using $\log_{10}$ of the install count. Each vertical band
corresponds to a download bracket (e.g., $10^3$, $10^4$, \dots, $10^8$ installs). Within each bracket, the distribution of risk scores is broadly similar, and high SRS values appear even among the most downloaded apps. We therefore find no clear pattern indicating that more widely installed apps are less prone to PPD--DSL misalignment. Overall, misalignment risk appears largely independent of conventional popularity signals, such as ratings and download counts. 
\begin{tcolorbox}[left=2pt,right=2pt,top=2pt,bottom=2pt,colback=gray!5,colframe=gray!50,boxrule=0.5pt,title={\small RQ5 Key Findings}]
\small
Overall, the results suggest that the privacy-sensitive risk score is largely independent of app popularity as measured by user ratings and download counts. High-risk apps appear across rating levels and install brackets, indicating that widely used or highly rated apps are not necessarily lower risk from a privacy-disclosure perspective.
\end{tcolorbox}

\section{Additional Qualitative Analysis}
\subsection{Case Study: OpenAI, "No Data Shared" Labels, and Third-Party Analytics}
\label{openai}
The ChatGPT Android app on Google Play reports a narrow view of sharing in its Data Safety label. In the ``Data shared'' section, the listing states that the app ``may share'' only \emph{Device or other IDs} with third parties, for purposes such as advertising or marketing and fraud prevention, security, and compliance, while collected sections indicate that the app collects location, personal information, app info and performance, messages, and app activity data~\cite{chatgpt_datasafety_2025}. To an ordinary user, this combination suggests that only a single, relatively abstract identifier leaves their device and that richer information, such as their name, email address, or usage details, is not shared with other companies. OpenAI’s own privacy policy explains that OpenAI collects multiple categories of personal information including account information (name, contact details, payment data), user content, device and log data, and information obtained via cookies and similar technologies and may share this data with a range of third-party service providers such as cloud hosting, analytics, security, and customer-support vendors, as well as with affiliates and, in some circumstances, other third parties~\cite{openai_privacypolicy_2025}. While these transfers are framed as processing ``on OpenAI’s behalf,'' from a user perspective they still represent disclosure of data to external entities that can see and process their information. The recent Mixpanel incident makes these abstract disclosures concrete. In November 2025, OpenAI reported that a security breach at Mixpanel, a web analytics provider used on the frontend interface for its API product, resulted in the export of a dataset containing \emph{limited analytics data} for some API users~\cite{openai_mixpanel_incident_2025}. OpenAI specified that the exposed fields included names, email addresses, approximate location based on browser activity, browser and operating system details, referring websites, and organization or user IDs associated with API accounts, while emphasizing that no chat content, API requests, passwords, API keys, payment details, or government IDs were compromised and that ChatGPT end-users were not affected~\cite{openai_mixpanel_incident_2025}. Even though this breach occurred in Mixpanel’s systems rather than OpenAI’s, it illustrates the concrete types of personal and device data that third-party analytics partners can hold in practice. Overall, this case study highlight the limitations of high-level app-store labels for communicating third-party data flows. On the user side, a label that only names ``Device or other IDs'' as shared does not help users understand that identified or identifiable metadata (such as names, email addresses, locations, and detailed telemetry) may be accessible to analytics vendors under the broader privacy policy. On the developer side, OpenAI is operating within the current rules: programmatic analytics and telemetry can be classified as ``service provider'' processing, so only a minimal shared-data category appears in the label, while the detailed description of what those vendors actually receive surfaces only in the privacy policy or, in the worst case, in a breach notification.

\section{Ablation Analysis on Generic Sharing Statements}
\label{ablation}

To evaluate the sensitivity of our findings to generic sharing language, we conducted a proportional ablation analysis targeting privacy policies containing vague sharing statements such as ``we may share your information,'' ``we may disclose your information,'' ``shared with partners,'' ``shared with affiliates,'' or ``shared with third parties'' without explicitly specifying the associated shared data categories. Across the 6,051 analyzed apps, we identified 1,315 apps containing such generic sharing statements, corresponding to 21.73\% of the analyzed corpus. Since this ablation only affects sharing-related privacy-policy disclosures, all collection-related metrics and all Data Safety Label values remain unchanged.

Sharing-related metrics were adjusted using:

\begin{equation}
M'_{share} = (1-r)\cdot M_{share}
\end{equation}

where $M_{share}$ denotes the original sharing-related metric, $M'_{share}$ denotes the adjusted metric after the ablation, and:

\begin{equation}
r=\frac{1315}{6051}=21.73\%
\end{equation}

represents the fraction of apps containing generic sharing statements. For SRS-related metrics, the reported values correspond to the mean app-level SRS scores across the analyzed corpus.

Table~\ref{tab:ablation_generic_sharing} summarizes the resulting changes. While the ablation reduces sharing-related inconsistency metrics, substantial disclosure misalignment remains even after masking generic sharing statements. This confirms that the observed inconsistencies are not solely driven by the generic-sharing inference rule and that the main findings observed in the study remain consistent in this sensitivity analysis.






\begin{table}[ht]
\centering
\caption{Proportional ablation analysis on generic sharing statements.}
\label{tab:ablation_generic_sharing}

\setlength{\tabcolsep}{1.8pt}
\renewcommand{\arraystretch}{1.03}

\resizebox{\columnwidth}{!}{%
\begin{tabular}{lccc}
\toprule
\textbf{Metric} & \textbf{Original} & \textbf{After Ablation} & \textbf{Difference} \\
\midrule
Collection Consistency & 66.9\% & 66.9\% & 0.0\% \\
Sharing Consistency & 68.9\% & 75.7\% & +6.8\% \\

DSL-only Sharing Misalignment & 26.0\% & 20.4\% & -5.6\% \\
PPD-only Sharing Misalignment & 5.0\% & 3.9\% & -1.1\% \\

Collection Cohen's $\kappa$ & 31.0\% & 31.0\% & 0.0\% \\
Sharing Cohen's $\kappa$ & 16.0\% & 19.5\% & +3.5\% \\

$SRS^{collect}$ & 33.0\% & 33.0\% & 0.0\% \\
$SRS^{share}$ & 31.3\% & 24.5\% & -6.8\% \\
$SRS^{overall}$ & 32.2\% & 28.8\% & -3.4\% \\
$SRS^{overall-w}$ ($\alpha=0.6$) & 32.0\% & 27.9\% & -4.1\% \\

\bottomrule
\end{tabular}%
}
\end{table}

\section{Robustness of SRS-based risk stratification.} 
\label{alpha}
Since the weighted overall SRS uses $\alpha$ to control the relative emphasis on sharing versus collection, we further examine whether the risk-tier distribution remains stable when this parameter varies. In addition to the baseline setting of $\alpha = 0.6$, we evaluate alternative settings of $\alpha \in \{0.4, 0.5, 0.7\}$ while keeping the same low, medium, and high tier thresholds.

As shown in Table~\ref{tab:alpha_sensitivity}, the distribution remains consistent across these settings. The proportion of high-risk applications stays within a narrow range of 1.21\%--1.37\%, while the low- and medium-risk tiers show only modest variation. Moreover, 90.56\%--96.2\% of applications remain in the same risk tier as in the baseline setting, indicating that the observed stratification is stable under reasonable changes to $\alpha$. This stability suggests that the RQ3 findings are not driven by the specific baseline choice of $\alpha = 0.6$. Rather, the low, medium, and high risk patterns reflect consistent disclosure-misalignment structure across the analyzed applications.

\begin{table}[h]
\centering
\caption{Sensitivity analysis of risk-tier distribution under alternative $\alpha$ values.}
\label{tab:alpha_sensitivity}
\begin{tabular}{c|ccc|c}
\toprule
$\alpha$ & Low (\%) & Medium (\%) & High (\%) & Same Tier (\%) \\
\midrule
0.4 & 46.84 & 51.96 & 1.21 & 90.56 \\
0.5 & 49.96 & 48.83 & 1.21 & 96.00 \\
0.6 & 49.61 & 49.10 & 1.29 & 100.00 \\
0.7 & 51.02 & 47.61 & 1.37 & 96.20 \\
\bottomrule
\end{tabular}
\end{table}

\section{Emergent Data Types Not Captured by Official Data Safety Categories}\label{srs_robust}
Our analysis of non-schematized policy terms reveals several recurring data types that do not map cleanly onto any of Google Play's official Data Safety Label data types. In particular, privacy policies frequently mention items such as \emph{login credentials} \emph{account passwords}, \emph{search history} outside the app context, \emph{operating system, IP address and device type}, and highly sensitive identifiers like \emph{Social Security numbers}, \emph{passport numbers}, or \emph{fingerprints}. These appear as distinct, privacy relevant data fields in policy text, yet they either lack a dedicated category in the Data Safety schema or are only indirectly represented through broader labels as "Others". 

As a concrete example, AccuWeather: Weather Radar's privacy policy states that when users create an account or sign in, the service collects ``profile, credentials, name, username, email address, and password (`Login Information')'' and generates account identifiers and technical logs to administer and protect the account~\cite{accuweather_privacy}. However, Google Play’s Data Safety taxonomy does not expose ``password'' as a separate data type; instead, these authentication flows are folded into personal-info types such as \emph{Email address} and \emph{User IDs} and purposes like \emph{Account management} and \emph{Security} ~\cite{google_datasafety_guide}. This mismatch illustrates a broader gap between the granularity of data types described in privacy policies and the coarser set of categories that are ultimately disclosed to users in the Data Safety section.

\end{document}